\documentclass[a4paper,fleqn]{cas-sc}
\usepackage[authoryear,longnamesfirst]{natbib}

\usepackage[T1]{fontenc} 
\usepackage{pdflscape} 
\usepackage{acro} 
\usepackage{hyperref} 
\usepackage[section]{placeins} 

\usepackage{tabularx} 
\usepackage{booktabs} 
\usepackage{caption} 
\usepackage{makecell} 
\usepackage{multirow} 
\usepackage{array} 
\usepackage{arydshln} 
\newcolumntype{P}[1]{>{\centering\arraybackslash}p{#1}} 
\newcolumntype{R}[1]{>{\raggedleft\arraybackslash}p{#1}} 

\usepackage{xcolor}

\DeclareAcronym{BTC}{
	short = BTC,
	long = Bitcoin
}
\DeclareAcronym{ETH}{
	short = ETH,
	long = Ethereum
}
\DeclareAcronym{LSTM}{
	short = LSTM,
	long = Long Short-Term Memory
}
\DeclareAcronym{NLP}{
	short = NLP,
	long = Natural Language Processing
}
\DeclareAcronym{AI}{
	short = AI,
	long = Artificial Intelligence
}
\DeclareAcronym{ML}{
	short = ML,
	long = Machine Learning
}
\DeclareAcronym{LLM}{
	short = LLM,
	long = Large Language Model
}
\DeclareAcronym{OLS}{
	short = OLS,
	long = Ordinary Least Squares
}
\DeclareAcronym{RNN}{
	short = RNN,
	long = Recurrent Neural Network
}
\DeclareAcronym{TFT}{
	short = TFT,
	long = Temporal Fusion Transformer
}
\DeclareAcronym{GRU}{
	short = GRU,
	long = Gated Recurrent Unit
}
\DeclareAcronym{GRN}{
	short = GRN,
	long = Gated Residual Network
}
\DeclareAcronym{SVM}{
	short = SVM,
	long = Support Vector Machine
}
\DeclareAcronym{CNN}{
	short = CNN,
	long = Convolutional Neural Network
}
\DeclareAcronym{FNN}{
	short = FNN,
	long = Feed-Forward Neural Network
}
\DeclareAcronym{MLP}{
	short = MLP,
	long = Multi-Layer Perceptron
}
\DeclareAcronym{AUC ROC}{
	short = AUC ROC,
	long = Area Under the Receiver Operating Characteristic Curve
}
\DeclareAcronym{MSE}{
	short = MSE,
	long = Mean Squared Error
}
\DeclareAcronym{BCE}{
	short = BCE,
	long = Binary Cross-Entropy
}
\DeclareAcronym{BPTT}{
	short = BPTT,
	long = Backpropagation Through Time
}
\DeclareAcronym{ARIMA}{
	short = ARIMA,
	long = Autoregressive Integrated Moving Average
}
\DeclareAcronym{MALSTM-FCN}{
	short = MALSTM-FCN,
	long = Multivariate Attention Long Short Term Memory Fully Convolutional Network
}

\begin{document}
\let\WriteBookmarks\relax
\def\floatpagepagefraction{1}
\def\textpagefraction{.001}

\shorttitle{Deep Learning and NLP in Cryptocurrency Forecasting: Integrating Financial, Blockchain, and Social Media Data}    

\shortauthors{V. Gurgul et al.}  

\title [mode = title]{Deep Learning and NLP in Cryptocurrency Forecasting: Integrating Financial, Blockchain, and Social Media Data}  

\author[1]{Vincent Gurgul}[orcid=0009-0002-1502-3670]
\cormark[1]
\ead{vincent.gurgul@hu-berlin.de}
\author[1, 2]{Stefan Lessman}[orcid=0000-0001-7685-262X]
\author[1]{Wolfgang Karl Härdle}[orcid=0000-0001-5600-3014]
\cortext[cor1]{Corresponding author}

\affiliation[1]{organization={Humboldt-Universität zu Berlin},
            addressline={Unter den Linden 6, 10117}, 
            city={Berlin},
            country={Germany}}
\affiliation[2]{organization={Bucharest University of Economic Studies},
            addressline={Piața Romană 8, 010374}, 
            city={Bucharest},
            country={Romania}}

\begin{abstract}
We introduce novel approaches to cryptocurrency price forecasting, leveraging \ac{ML} and \ac{NLP} techniques, with a focus on Bitcoin and Ethereum. By analysing news and social media content, primarily from Twitter and Reddit, we assess the impact of public sentiment on cryptocurrency markets. A distinctive feature of our methodology is the application of the BART MNLI zero-shot classification model to detect bullish and bearish trends, significantly advancing beyond traditional sentiment analysis. Additionally, we systematically compare a range of pre-trained and fine-tuned deep learning NLP models against conventional dictionary-based sentiment analysis methods. Another key contribution of our work is the adoption of local extrema alongside daily price movements as predictive targets, reducing trading frequency and portfolio volatility. Our findings demonstrate that integrating textual data into cryptocurrency price forecasting not only improves forecasting accuracy but also consistently enhances the profitability and Sharpe ratio across various validation scenarios, particularly when applying deep learning NLP techniques. The entire codebase of our experiments is made available via an online repository: \url{https://anonymous.4open.science/r/crypto-forecasting-public}
\end{abstract}



\maketitle


\section{Introduction}

The cryptocurrency market has emerged as a digital economy over the last decade, attracting substantial attention from researchers and practitioners. This unique decentralised market is characterised by high volatility and ample data availability, making it a compelling field for the application of \ac{AI} and \ac{ML} techniques. In particular, the availability of vast public sentiment data, primarily from social networks, opens up new avenues for the integration of \ac{NLP} into the forecasting of cryptocurrency price movements.

The research presented in this work examines the impact of news from various cryptocurrency-related outlets and social media posts from Twitter and Reddit on the valuations of \ac{BTC} and \ac{ETH}, the two largest cryptocurrencies by market capitalisation. Traditionally, studies in the cryptocurrency domain have leaned on dictionary-based methods to analyse the influence of news and social media. However, with the advancements in linguistic \ac{AI} models, there is an opportunity to explore new avenues for sentiment analysis. Our research extends beyond traditional techniques by incorporating deep learning \ac{NLP} methods to gauge market sentiment. While the application of deep learning in sentiment analysis is established, our work distinguishes itself by adopting a zero-shot classification language model specifically to differentiate between `bullish' and `bearish' market perspectives, a nuanced approach that moves beyond general sentiment classification to provide more targeted insights into market dynamics.

We also advance beyond established practices in our price forecasting methodology. In the realm of cryptocurrency analysis, this is usually regression onto price changes or the binary classification of daily price movement (upward or downward). Local extrema remain unexplored as a target variable, even though their inherent lower noise levels offer a significant advantage in the highly volatile cryptocurrency market. We conjecture that by classifying local minima and maxima our \ac{ML} models obtain enhanced prediction performance, both with respect to classification metrics and profitability. In addition to the forecasting of daily price movements, we therefore aim to establish whether the textual data can aid in the prediction of local extreme points with various observational time frames.

We compare the predictive performance of multiple \ac{ML} methods---including deep learning and sequential models---using different approaches for quantifying market sentiment, to models that do not include textual data. This comparison is conducted across five different target variables and includes a trading simulation.

By exploring the evolution of market efficiency, our research traces the adaptation of markets to the increasing prominence of social media discourse over time, all the way since 2012. Our comprehensive approach, which also includes on-chain data, GitHub and Google Trends, enables us to reveal the nuanced ways in which information dissemination affects cryptocurrency market behaviour.


\section{Literature Review}\label{sec:literature}

Since 2015, there has been sustained interest in leveraging information from social media for forecasting cryptocurrency prices. This growing body of research has even led to the publication of review papers on the subject. To provide a comprehensive overview of the literature, we summarise the key findings from these reviews and also conduct an analysis of significant individual studies. In the latter we include papers that predict cryptocurrency prices or associated target variables using NLP data. Studies that do not make use of textual data in their modelling pipeline or that solely focus on forecasting volatility or examining statistical properties of the time series, such as change points, are not taken into account.

In the following subsections we dissect the forecasting methods, the NLP approaches, the target variable selection, and the variety of explanatory variables considered---aiming to establish a thorough understanding of how the field has developed and to identify prevailing trends and avenues for further exploration. The results are summarised in Table \ref{tab:crypto_forecasting_approaches}.

\subsection{Evolution of Forecasting Techniques}

The forecasting methods are categorized into linear, non-linear, and sequential models. Linear models, such as Exponential Smoothing, Autoregressive Moving Averages, \ac{OLS}, and Support Vector Machines are valued for their simplicity and interpretability. According to the review by \cite{fang_2022_cryptocurrency} OLS Regression has not only formed the backbone of cryptocurrency forecasting in its beginnings, but is still the most used forecasting method in the literature.

However, the landscape of financial analytics evolved, gradually shifting towards more sophisticated methods \citep{zhang_2024_a}. Non-linear machine learning approaches are capable of capturing more intricate interactions between the variables, yet they are less interpretable due to their complex internal structures. Multiple studies systematically evaluate various non-linear model types, such as Random Forests, Adaboost, Gradient Boosting, and \acp{MLP} in the context of cryptocurrency forecasting. These studies frequently highlight the superior performance of Gradient Boosting and \ac{MLP} neural networks \citep{valencia_2019_price, wok_2019_advanced}.

Sequential models, such as \acp{RNN} and Transformers, excel in handling ordered time-series data prevalent in financial contexts. The review by \cite{zhang_2024_a} reveals that many state-of-the-art neural network architectures are applied to cryptocurrencies, ranging from \ac{LSTM} and \ac{GRU} networks to \acp{CNN}. Yet, only 14 \% of cryptocurrency forecasting papers make use of neural networks at all, with sequential models accounting for only 4 \% \citep{fang_2022_cryptocurrency}. \cite{pant_2018_recurrent}, \cite{mittal_2019_shortterm}, \cite{raju_2020_realtime}, and \cite{kim_2023_cbits}, achieve very good results with standard \acp{RNN} and \ac{LSTM} networks. \cite{ortu_2022_on} additionally implement \acp{CNN} and \acp{MALSTM-FCN}. \cite{parekh_2022_dlguess} build an \ac{RNN}-based model that integrates sequential and non-sequential data side by side.

Even less widespread than standard \acp{RNN} is the application of Transformer-based time series models like the \ac{TFT}, with \cite{murray_2023_on} representing the only exploration to date. Our study aims to bridge this gap by further investigating the \ac{TFT} alongside \ac{OLS} Regression, Gradient Boosting, \acp{MLP}, and \ac{LSTM} networks.

\shortcites{colianni_2015_algorithmic}
\shortcites{abraham_2018_cryptocurrency}
\shortcites{jain_2018_forecasting}
\shortcites{karalevicius_2018_using}
\shortcites{pant_2018_recurrent}
\shortcites{chen_2019_what}
\shortcites{hao_2019_predicting}
\shortcites{inamdar_2019_predicting}
\shortcites{li_2019_sentimentbased}
\shortcites{mittal_2019_shortterm}
\shortcites{valencia_2019_price}
\shortcites{wok_2019_advanced}
\shortcites{bakar_2020_weighted}
\shortcites{derbentsev_2020_forecasting}
\shortcites{jay_2020_stochastic}
\shortcites{mudassir_2020_timeseries}
\shortcites{pintelas_2020_investigating} 
\shortcites{raju_2020_realtime}
\shortcites{livieris_2021_an}
\shortcites{ider_2022_cryptocurrency}
\shortcites{kim_2022_a}
\shortcites{ortu_2022_on}
\shortcites{parekh_2022_dlguess}
\shortcites{kim_2023_cbits}
\shortcites{murray_2023_on}
\shortcites{rafi_2023_enhancing}
\shortcites{subramanian_2024_a}

\begin{landscape}

	\begin{table}[H]
		\scriptsize
		\setlength\extrarowheight{1.8pt}
		\centering
		\caption{Overview of cryptocurrency price forecasting approaches that utilise NLP}
		\vspace{-0.2em}
		\begin{tabular}{|l|*{2}{P{2.1em}:}P{2.1em}|*{3}{P{1.5em}:}P{1.5em}|*{2}{P{1.5em}:}P{1.5em}|*{6}{P{1.5em}:}P{1.5em}|c|}
			\hline
			\multirow{2}{*}{{Paper}} &
			\multicolumn{3}{c|}{{Forecasting Methods}} &
			\multicolumn{4}{c|}{{NLP Methods}} &
			\multicolumn{3}{c|}{{Targets}} &
			\multicolumn{7}{c|}{{Features}} &
			\multirow{2}{*}{\rotatebox[origin=c]{90}{\smash{\makebox[18.1ex][l]{\hspace{-1.6em}Cryptocurrencies$^1$}}}} \\
			& \rotatebox[origin=c]{90}{Linear Models} & \rotatebox[origin=c]{90}{~\,Non-linear Models~} & \rotatebox[origin=c]{90}{ Sequential Models} & \rotatebox[origin=c]{90}{Dictionary} & \rotatebox[origin=c]{90}{Embeddings} & \rotatebox[origin=c]{90}{RNN} & \rotatebox[origin=c]{90}{Transformers} & \rotatebox[origin=c]{90}{Regression} & \rotatebox[origin=c]{90}{Classification} & \rotatebox[origin=c]{90}{Local extrema} & \rotatebox[origin=c]{90}{Financial} & \rotatebox[origin=c]{90}{Blockchain} & \rotatebox[origin=c]{90}{News} & \rotatebox[origin=c]{90}{Twitter} & \rotatebox[origin=c]{90}{Reddit} & \rotatebox[origin=c]{90}{GitHub} & \rotatebox[origin=c]{90}{Google Trends} & \\
			\hline
			\cite{colianni_2015_algorithmic} & \checkmark & \checkmark & & \checkmark & & & & & \checkmark & & & & & \checkmark & & & & b \\
			\hline
			\cite{abraham_2018_cryptocurrency} & \checkmark & & & \checkmark & & & & \checkmark & & & & & & \checkmark & & & \checkmark & be \\
			\hline
			\cite{jain_2018_forecasting} & \checkmark & & & \checkmark & & & & \checkmark & & & & & & \checkmark & & & & bl \\
			\hline
			\cite{karalevicius_2018_using} & & & & \checkmark & & & & & \checkmark & & & & \checkmark & & & & & b \\
			\hline
			\cite{pant_2018_recurrent} & & & \checkmark & & \checkmark & & & \checkmark & & & & & & \checkmark & & & & b \\
			\hline
			\cite{chen_2019_what} & \checkmark & & & \checkmark & & & & \checkmark & & & & & & \,\checkmark$^2$ & \checkmark & & & i \\
			\hline
			\cite{hao_2019_predicting} & & \checkmark & & \checkmark & & & & \checkmark & & & & & & \checkmark & & & \checkmark & b \\
			\hline
			\cite{inamdar_2019_predicting} & & \checkmark & & & & \checkmark & & \checkmark & & & & & \checkmark & \checkmark & & & & b \\
			\hline
			\cite{li_2019_sentimentbased} & & \checkmark & & \checkmark & & & & \checkmark & & & \checkmark & & & \checkmark & & & & z \\
			\hline
			\cite{mittal_2019_shortterm} & \checkmark & \checkmark & \checkmark & \checkmark & & & & \checkmark & & & & \checkmark & & \checkmark & & & \checkmark & b \\
			\hline
			\cite{valencia_2019_price} & \checkmark & \checkmark & & \checkmark & & & & & \checkmark & & \checkmark & & & \checkmark & & & & berl \\
			\hline
			\cite{wok_2019_advanced} & \checkmark & \checkmark & & \checkmark & & & & \checkmark & & & \checkmark & \checkmark & & \checkmark & & & \checkmark & bermnc \\
			\hline
			\cite{raju_2020_realtime} & \checkmark & & \checkmark & \checkmark & & & & \checkmark & & & \checkmark & & & \checkmark & & & & b \\
			\hline
			\cite{ider_2022_cryptocurrency} & \checkmark & \checkmark & & & & & \checkmark & \checkmark & \checkmark & & \checkmark & \checkmark & \checkmark & \checkmark & \checkmark & & & be \\
			\hline
			\cite{ortu_2022_on} & & \checkmark & \checkmark & & & & \checkmark & & \checkmark & & \checkmark & & & & \checkmark & \checkmark & & be \\
			\hline
			\cite{parekh_2022_dlguess} & & & \checkmark & \checkmark & & & & \checkmark & & & & & & \checkmark & & & & bld \\
			\hline
			\cite{kim_2023_cbits} & & \checkmark & \checkmark & & & & \checkmark & & \checkmark & & \checkmark & & \checkmark & & & & & b \\
			\hline
			\cite{subramanian_2024_a} & \checkmark & & \checkmark & \checkmark & & & & \checkmark & & & & & \checkmark & \checkmark & & & & b \\
			\hline
			The proposed approach & \checkmark & \checkmark & \checkmark & \checkmark & & & \checkmark & \checkmark & \checkmark & \checkmark & \checkmark & \checkmark & \checkmark & \checkmark & \checkmark & \checkmark & \checkmark & be \\
			\hline
        \end{tabular}
        \\\vspace{0.5em}
        \tiny{
			$^1$ b = Bitcoin, e = Ethereum, l = Litecoin, m = Monero, z = ZClassic, n = Electroneum, c = ZCash, d = Dash, i = CRIX index
			\vspace{0.3em}\\
			$^2$ \cite{chen_2019_what} leverage the investment-specific platform StockTwits instead of Twitter
		}
        \label{tab:crypto_forecasting_approaches}
    \end{table}

\end{landscape}

\subsection{Application of \ac{NLP} Methods}

\ac{NLP} methods are crucial for understanding market sentiment, a significant driver of cryptocurrency price fluctuations. The review by \cite{fang_2022_cryptocurrency} determines two main approaches to utilising textual data in this context: 1) assigning a sentiment score to unlabelled text data using an unsupervised method, such as a sentiment dictionary or a pre-trained Transformer, and 2) labelling the text data (either manually or using the price movements, assuming that a price increase implies a positive market sentiment) and subsequently taking a supervised learning approach to training a machine learning model, such as an \ac{RNN}. However, there are also intermediate approaches that do not fit neatly into either category, for example, using pre-trained (unsupervised) embeddings and feeding the results into a supervised numerical model, or fine-tuning a pre-trained Transformer. Instead of the training approach, we therefore divide the literature based on the language model architectures, which can be classified into dictionary-based approaches, embeddings, \acp{RNN}, and Transformers.

Dictionary-based approaches assign sentiment scores to words based on pre-defined, often manually labelled dictionaries. While computationally efficient, they fail to capture long-term dependencies and contextual nuances \citep{liu_2012_sentiment}. According to \cite{fang_2022_cryptocurrency}, dictionary-based sentiment analysis is highly prevalent in cryptocurrency price forecasting with the most commonly employed dictionaries being the general-purpose emotional valence libraries VADER and Textblob. In addition to sentiment scores for each word, those two consider a list of `negators' and `intensifiers' (words that reverse or amplify/weaken sentiment). \cite{valencia_2019_price} report significant improvements in Bitcoin and Litecoin price forecasting accuracy when including VADER sentiment scores from Twitter data. Some studies make use of domain-specific lexicons, such as \cite{karalevicius_2018_using}, who compare the Harvard Psychosocial and Loughran-McDonald finance-specific dictionaries, and \cite{chen_2019_what}, who develop a cryptocurrency-specific lexicon using posts from the StockTwits platform, where users label their posts as `bullish' or `bearish' themselves.

Embeddings offer a more sophisticated mechanism for capturing semantic meanings by encoding words as high-dimensional vectors \citep{mikolov_2013_distributed}. Those vectors can then be fed as a numeric input into a standard machine learning model. \cite{pant_2018_recurrent} leverage the pre-trained Gensim Word2Vec embeddings alongside a Bag-of-Words approach to include textual data from Twitter in their cryptocurrency forecasting pipeline.

\acp{RNN} have the capability to model contextual relationships over sequences, making them suitable for capturing dependencies in text \citep{graves_2012_supervised}. \cite{inamdar_2019_predicting} train an \ac{LSTM} network on tweets and news articles labelled with historical Bitcoin price data and subsequently feed those features into a Random Forest model for future price prediction.

Transformer architectures represent the cutting edge of \ac{NLP} by excelling in capturing complex long-range dependencies in language, which enables them to model intricate grammatical and logical patterns \citep{vaswani_2017_attention}. \cite{ortu_2022_on} employ a pre-trained BERT model to categorize emotions and sentiments in cryptocurrency-related comments on GitHub and Reddit, analysing both sentiment polarity and specific emotional reactions, such as love, joy, anger, or sadness. They report a significant improvement in forecasting accuracy on hourly data when including the sentiment scores. \cite{kim_2023_cbits} fine-tune BERT, DeBERTa, and RoBERTa models on a manually labelled cryptocurrency-specific corpus of news articles and compare their performance to pre-trained mBERT and XLM-RoBERTa models. They show that the fine-tuned models slightly outperform the pre-trained models, with the fine-tuned RoBERTa taking the lead. Furthermore, they state that including NLP data resulted in a 3 \% increase in accuracy and 20 \% increase in profits with no change in volatily.

Outside of price forecasting, there exists a broader literature on advanced sentiment analysis methods in the realm of cryptocurrencies \citep{widianto_2023_sentiment, dwivedi_2023_cryptocurrency}, but their potential for improving predictive modelling remains underexplored. We aim to address this gap by further exploring Transformer models, fine-tuning them, and introducing a novel approach for converting textual data into numerical representations of market sentiment. The VADER dictionary is included as a benchmark due to its proven effectiveness in previous research.

\subsection{Target and Feature Selection}

Literature predominantly treats price forecasting as a regression problem, predicting the next period's price or its relative change \citep{fang_2022_cryptocurrency}. Classification approaches, which predict price increases or decreases, are less common but have demonstrated superior performance both in terms of classification metrics and trading profit \citep{leung_2000_forecasting, mudassir_2020_timeseries}. The prediction of local extreme points, however, remains unexplored in cryptocurrencies. Recognising the potential of discrete targets, our research expands beyond price regression, incorporating classification and, distinctively, the forecasting of local extrema.

\cite{fang_2022_cryptocurrency} observe a diverse range of predictors, ranging from financial data (indicators and other assets) to blockchain data (information contained on the cryptocurrencies' ledgers, e.g.\ transaction and balance records), textual data (news outlets, Twitter, Reddit), GitHub data, and Google Trends. Initially, research focused on autoregressive analyses and sentiment data from Twitter and news outlets. Early on, Google Trends, which refers to the volume of searches for specific related keywords on Google, are integrated as a feature in various approaches. \cite{fang_2022_cryptocurrency} highlight that many papers report a significant relationship between Google Trends and cryptocurrency price movements. More recently, however, there is a trend towards using blockchain metrics, transaction data, and Reddit discussions. Our analysis employs a comprehensive set of explanatory variables to evaluate the relative merit of these diverse data sources.


\section{Methodology}\label{sec:methods}

Building on the comprehensive review of prior literature, this section delves into the methodological framework of our study. We begin by exploring the realm of time series modelling through the lens of neural networks, shedding light on their respective utilities and the advancements they have introduced in modelling financial time series data. Subsequently, we present an extensive overview of neural network-based \ac{NLP} methods, beginning with embeddings and advancing to more sophisticated techniques such as \acp{RNN} and Transformer models. In this context, we introduce and explain the three deep learning models at the forefront of our textual analysis: (i) Twitter-RoBERTa, a sentiment analysis model trained specifically on social media data; (ii) BART MNLI, a zero-shot classification model that we tailor for gauging `bullishness' in financial narratives; and (iii) a vanilla RoBERTa model, fine-tuned on our targets. Finally, our discourse shifts to the rationale behind our target variable selection and our trading strategy. Here, we elaborate on our choice of targets, detail the process of creating local extreme point targets, and articulate the principles that shape our approach to market entry and exit.

\subsection{Time Series Modelling with Neural Networks}\label{sec:tsm}

Traditional statistical methods, such as \ac{ARIMA} and Exponential Smoothing, have been the go-to time series modelling techniques for decades. However, with the advent of deep learning, neural networks have emerged as a powerful alternative, offering the potential to capture more complex patterns and relationships in time series data \citep{zhang_1998_forecasting}. This has led research in the realm of financial analysis to increasingly turn to deep learning methodologies for price forecasting \citep{ozbayoglu_2020_deep}. The following section provides a concise overview of neural-network-based time series modelling approaches, followed by a comprehensive examination of the \ac{TFT} architecture.

The simplest form of a neural network is referred to as an \ac{MLP} or \ac{FNN}. It comprises multiple layers of nodes, commonly known as neurons. Specifically, there is an input layer, one or more hidden layers, and an output layer. The strength of the connections between the nodes is defined by weights and biases, which are both adjusted using backpropagation. During backpropagation, the error of each prediction on the training dataset is circulated backward through the network, determining the contribution of each node to the overall discrepancy. By leveraging a differentiable objective function and an optimisation algorithm, such as gradient descent, the weights and biases are iteratively tuned to minimise the error. In the case of time series data, past observations of the explanatory time series serve as input features, while future values of the time series of interest are used as the target for error computation \citep[for an in-depth explanation of neural networks see][]{goodfellow_2016_deep}.

\begin{figure}[!h]
	\centering
	\includegraphics[width=0.65\textwidth]{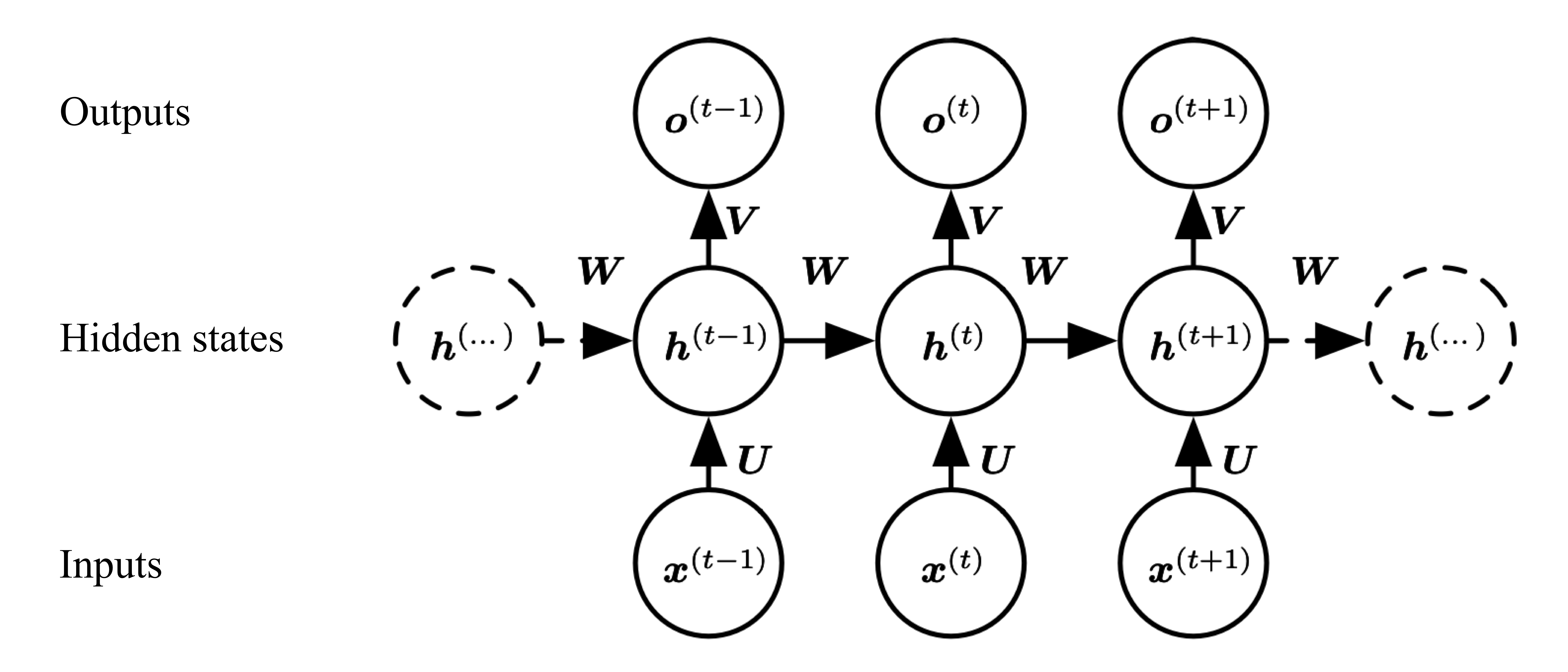}
	\\\vspace{0.8em}
	\small{$U,~V,~W:\,$ model weight matrices\\~~~$t:\,$ present timestep}
	\\\vspace{1.2em}
	\caption{The standard \ac{RNN} model architecture \citep[Adapted from][]{goodfellow_2016_deep}}
	\label{fig:rnn}
\end{figure}

Among the various neural network architectures, \acp{RNN} have gained prominence due to their inherent design tailored for sequential data, that is, data ordered based on the occurrence of events over the course of time. While \acp{FNN} consider each of the lagged features independently, an \ac{RNN} operates in a more temporally aware fashion. Specifically, an \ac{RNN} bases its predictions on the most recent lag and its internal hidden state, which encapsulates historical lags and their intricate relationships. Conversely, \acp{FNN}, while utilising all provided lags, neglect the inherent sequential order and potential temporal patterns present in the data. When referring to an \ac{RNN} with $n$ neurons, we typically mean that the inputs are encoded as $n$-dimensional hidden states. Those hidden states can then be processed by passing them through an \ac{FNN} to produce the desired output shape or by iterating the \ac{RNN} architecture to generate forecasts multiple timesteps into the future.

\acp{RNN} are termed `recurrent' because their previous outputs---the past hidden states---are fed back into future iterations as input. However, \acp{RNN} are not recurrent in the sense that information ever loops back from a neuron to itself; it also passes linearly from the input to the output layer. Any \ac{RNN} can, therefore, be unfolded in time and represented as an \ac{FNN} for a fixed-length sequence. This representation enables the application of gradient-based techniques to train the network, a concept known as Backpropagation Through Time \citep[see further details on \acp{RNN} in][]{goodfellow_2016_deep}.

\begin{figure}[!h]
	\centering
	\includegraphics[width=0.6\textwidth]{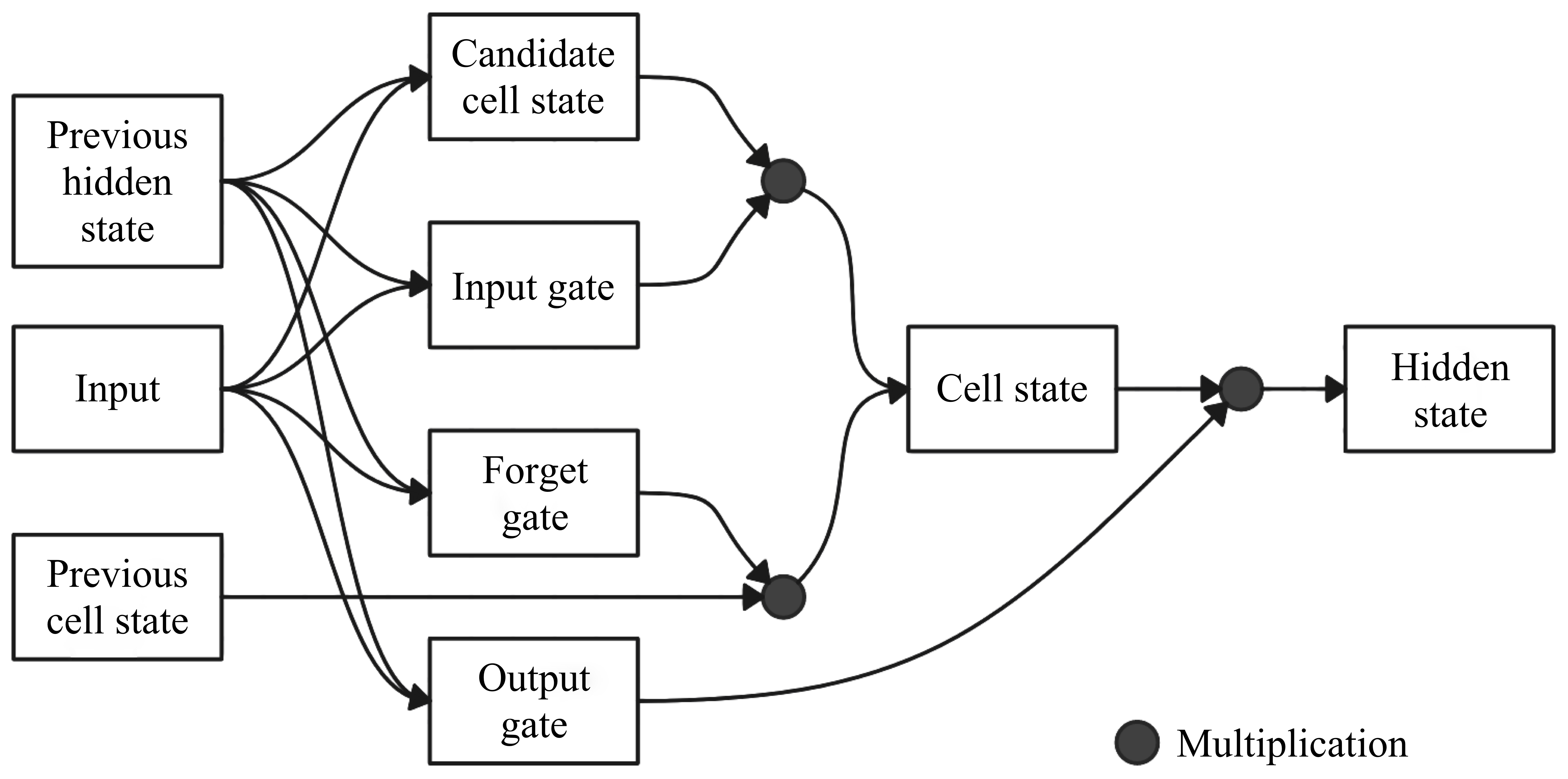}
	\vspace{1.5em}
	\caption{\ac{LSTM} cell architecture}
	\label{fig:lstm}
\end{figure}

Advanced \ac{RNN} architectures, notably \ac{LSTM} and \ac{GRU} networks, have emerged to address some intrinsic limitations of vanilla \acp{RNN}. One such challenge is the vanishing gradient problem, where the gradients diminish significantly across successive backpropagation steps, impeding the network's capacity to learn from distant past events. The \ac{LSTM} architecture, first introduced  by \cite{hochreiter_1997_long} and then refined by \cite{gers_2000_learning}, combats this issue through a more intricate design. In addition to a hidden state, it possesses a cell state, which serves as the memory of the network and can carry information untouched through many timesteps, complemented by three gates---input, forget, and output---that control the information flow.

\begin{figure}[!h]
	\centering
	\includegraphics[width=0.9\textwidth]{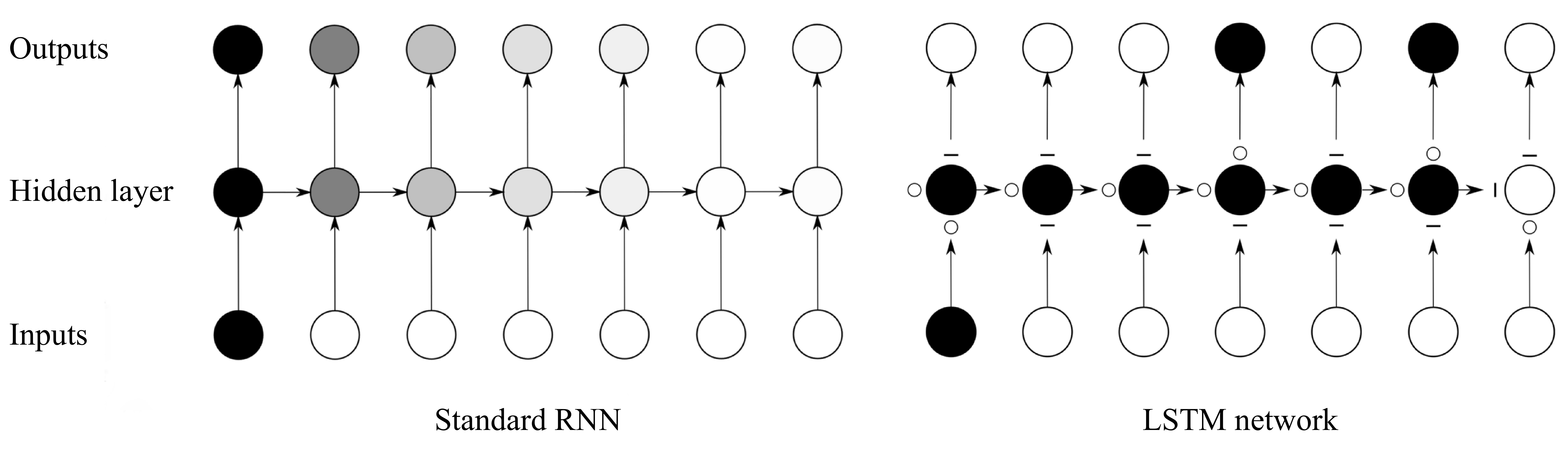}
	\vspace{0.4em}
	\caption{Comparison of information decay in \acp{RNN} and \ac{LSTM} networks \citep[Adapted from][]{graves_2012_supervised}}
	\label{fig:gradient_decay}
\end{figure}

Fig. \ref{fig:gradient_decay} illustrates the vanishing gradient problem and how the \ac{LSTM} addresses it. In the standard \ac{RNN}, the information decays over time as new inputs overwrite the hidden state. The \ac{LSTM} cell state, on the other hand, passes the information from the first input as long as the forget gate is open and the input gate is closed. In practice, the gates are not strictly binary; instead, they can take on any continuous value between zero and one, allowing for nuanced modulation of the information flow.

\acp{GRU} offer a streamlined alternative to \acp{LSTM} by consolidating the cell and hidden states and employing just two gates---the update and reset gates. Despite their simplified architecture, they often match \acp{LSTM} in performance while being more computationally efficient \citep[for a comprehensive overview of advanced \ac{RNN} architectures see][]{graves_2012_supervised}.

Another neural network architecture that is applicable in time series analysis is the \ac{CNN}. Unlike their primary use in image processing, where they identify spatial patterns, in time series, one-dimensional \acp{CNN} can identify and extract localised, shift-invariant patterns from the input data. The convolutional layers use sliding filters to learn temporal patterns, such as spikes, drops or specific shapes, with subsequent pooling and dense layers extracting dominant features and making predictions \citep[see][]{ismailfawaz_2019_deep}.

\begin{figure}[!h]
	\centering
	\includegraphics[width=0.83\textwidth]{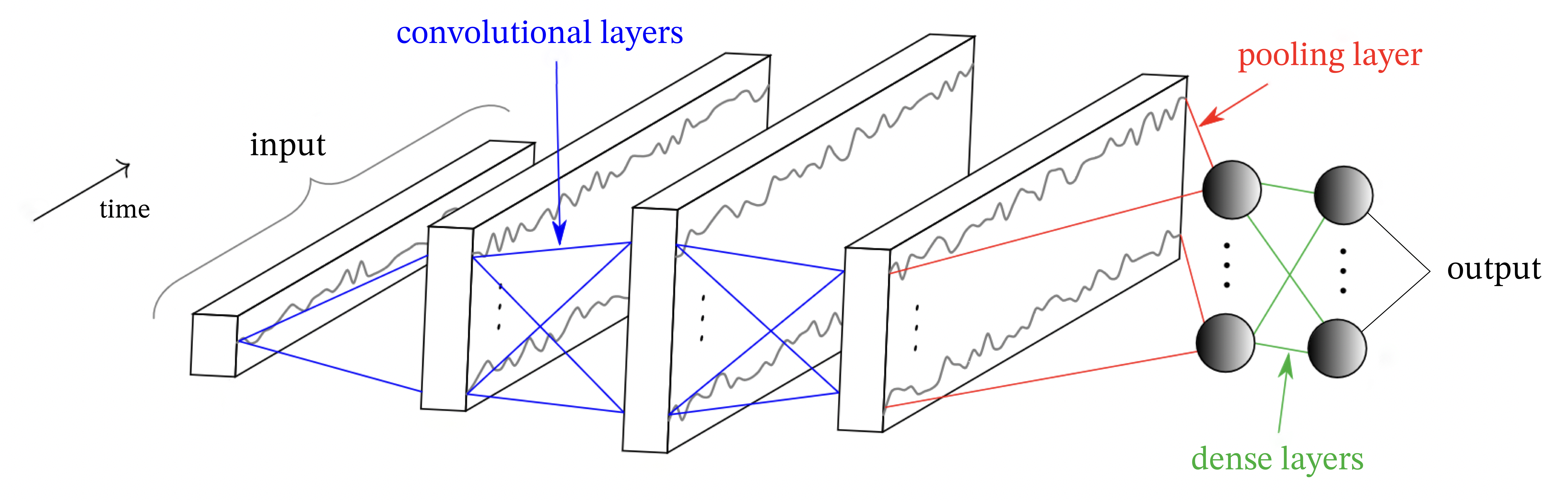}
	\vspace{0.4em}
	\caption{\ac{CNN} time series model architecture \citep[Adapted from][]{ismailfawaz_2019_deep}}
	\label{fig:cnn}
\end{figure}

The \ac{TFT}, introduced by \cite{lim_2020_temporal}, is a notable evolution in time series modelling, blending aspects of recurrent structures with attention mechanisms. Conceptually, the \ac{TFT} is best understood as a panel data model. Its design allows for the simultaneous modelling of individual-specific static metadata alongside the temporal component. This facilitates the simultaneous analysis of multiple entities across time, harnessing the full potential of both cross-sectional and time series data.

At the heart of the \ac{TFT} architecture lies an \ac{LSTM} encoder-decoder framework. This setup provides the model with the capability to transform input sequences into a compressed representation, which is subsequently decoded to produce the forecasted values. \acp{LSTM}, as detailed earlier, are adept at capturing sequential patterns and ensuring long-term dependencies are considered.

A distinguishing feature of the TFT is its incorporation of the Temporal Self-Attention mechanism. Borrowing insights from the Self-Attention of the Transformer language model, this mechanism enables the \ac{TFT} to reweigh different points in the input sequence. By doing so, the model can discern long-spanning seasonal patterns and dependencies, particularly in situations where the influence of past events is not uniform, but varies based on context.

\begin{figure}[!h]
	\centering
	\includegraphics[width=0.85\textwidth]{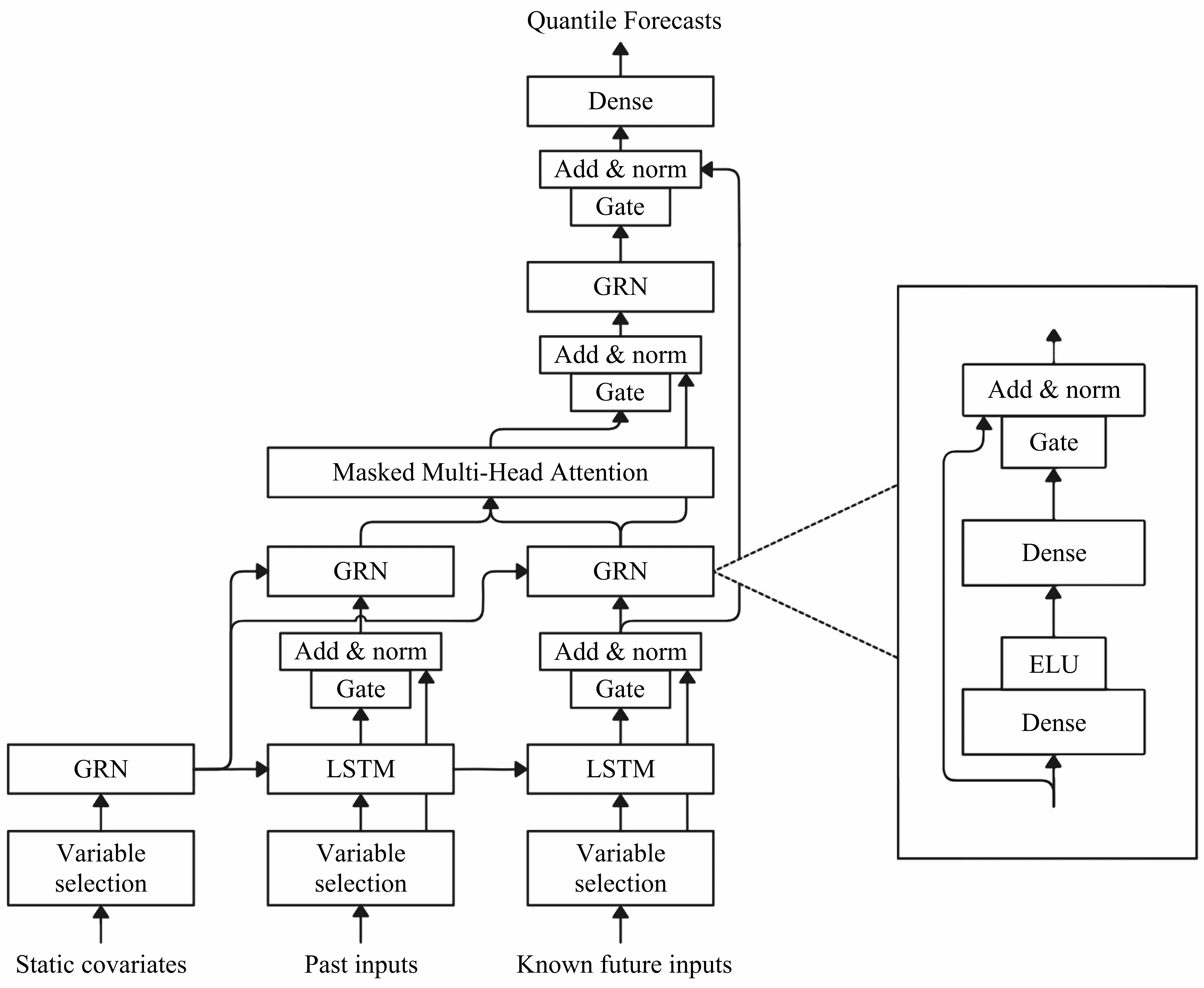}
	\vspace{0.8em}
	\caption{\ac{TFT} model architecture}
	\label{fig:tft}
\end{figure}

Integral to the \ac{TFT}'s design is its use of \acp{GRN} for variable selection and in the later layers of the network. The potential of a \ac{GRN} hinges on two key components. Firstly, the residual connection facilitates gradient propagation, allowing for the training of deeper networks without encountering the vanishing gradient problem. Secondly, the gating mechanism enables the \ac{GRN} to dynamically alternate between the original input and the transformed input. This adaptability helps in selectively emphasising certain features, proving useful for variable selection and facilitating the extraction of feature importances.

Yet, the merits of the \ac{TFT} do not end with the capability of modelling more complex temporal relationships. Two benefits elevate its utility in practical applications: the generation of quantile forecasts and the model's inherent explainability. Instead of a single point prediction, quantile forecasts provide a range of values, akin to a confidence interval, enabling more informed decision-making strategies. The explainability through feature importances and attention weights, ensures that despite the model's complexity, the significant drivers of its predictions are discernible. In contexts where understanding the rationale behind predictions can be as critical as the forecasts themselves, those attributes contribute to the \ac{TFT}'s appeal.

Our research delves into the application of sequential deep learning time series models for the forecasting of cryptocurrency prices, alongside traditional non-sequential models. The complex dynamics among the explanatory time series, particulary in relation to the local extreme points, may demand a more flexible model like the \ac{LSTM} network or the \ac{TFT}. The \ac{TFT} stands out as a prime candidate given its automatic variable selection and inherent explainability. With regard to sequential models, we also apply an \ac{LSTM} network, as it has consistently demonstrated superior performance over other \acp{RNN} on a variety of different datasets \citep{greff_2017_lstm}. A detailed overview of all employed \ac{ML} models is found in Section \ref{sec:models}.

\subsection{Deep Learning Approaches for \ac{NLP}}\label{sec:nlp_methods}

Deep learning methods have revolutionised the field of \ac{NLP}, offering context understanding, handling of linguistic ambiguities, reduced necessity for manual feature engineering, and improved generalisation capabilities. In doing so, they have enabled breakthroughs in end-to-end learning, transfer learning, multimodal integration and multilingual processing \citep{young_2017_recent, ruder_2019_a}.

In deep learning, words are represented as high-dimensional vectors, called embeddings, which are capable of capturing semantic and syntactic similarities \citep{mikolov_2013_distributed}. This is done through approaches like Word2Vec or GloVe, that learn from raw text, or through backpropagating loss of Transformer models like BERT, when trained on a specific task. These word embeddings are subsequently fed into an \ac{RNN}, \ac{CNN}, or an attention-based neural network like the Transformer \citep[for an in-depth overview of neural network architectures for \ac{NLP} see][]{goldberg_2017_neural}. These models aim to capture context by modelling long-term dependencies between words. This approach addresses language ambiguity, that is, the fact that the same word can have multiple meanings. Ultimately, this enables deep learning models to encode the meaning of a sentence, or even an entire piece of text, in a context-aware fashion \citep{reimers_2019_sentencebert}.

Deep learning models reduce the need for extensive feature engineering (like part-of-speech tagging or named entity recognition), a common requirement in traditional \ac{NLP}. Furthermore, they can learn useful features from raw text, removing the need for hand-labelled dictionaries and thus making them more scalable. This allows for end-to-end learning, where a single model processes raw text and directly outputs the final task results, such as classifications or translations, eliminating the need for complex multi-step pipelines common in traditional \ac{NLP} \citep{bahdanau_2014_neural, lecun_2015_deep}.

Furthermore, deep learning models have demonstrated significant efficacy in transfer learning applications within the realm of \ac{NLP}. \acp{LLM} like BERT or GPT, which are pre-trained on vast corpora, can be fine-tuned on specific tasks with relatively small datasets, leveraging knowledge learned from the large-scale text collections. The models first train on a corpus of unstructured and unlabeled text data, for example, by trying to predict the next word in a sequence. This allows early layers to extract general language features, such as syntax rules or semantic relationships and acts as a basic language understanding. During fine-tuning, this pre-trained model is adjusted to perform a specific task, like sentiment analysis. The early layers, already skilled in general language understanding, remain largely unchanged, while the later layers (e.g.\ a classification head) adapt to map the general language features to the specific task \citep{howard_2018_universal}.

In this work, we apply three different deep-learning-based \acp{LLM}. The first is the fine-tuned sentiment analysis model Twitter-RoBERTa-Base (version from 25.01.2023). It consists of an encoder from a Transformer model \citep{vaswani_2017_attention}, which was first pre-trained on 161 GB of raw text data to become RoBERTa-Base \citep{liu_2019_roberta}, and then fine-tuned for sentiment analysis on a manually labelled dataset of 124 million tweets \citep{loureiro_2022_timelms}. With this volume of training data, it stands out as the most exhaustive sentiment analysis model tailored for social media posts. The labels are positive, neutral, and negative, which we merge into a single sentiment score.

The second model is the fine-tuned zero-shot classifier BART-Large MNLI \citep{lewis_2019_bart}. This model utilises the encoder of a pre-trained BART-Large model and is fine-tuned on the MultiNLI dataset, which contains 433 thousand sentence pairs annotated with textual entailment information. Each data point consists of (i) a `premise'---a specific piece of text, (ii) a `hypothesis' that may or may not refer to this piece of text, and (iii) a label that indicates whether the hypothesis is true, false, or unrelated to the premise. For our methodology, we input our textual data into the model as the premise. As the hypothesis, we use the sentence ``This example is bullish for Bitcoin.'' or its Ethereum equivalent. The model then produces a score that reflects the probability of this hypothesis being true. This application of a zero-shot classification language model goes beyond what has been applied in existing financial forecasting literature.

Another contribution is the further exploration of fine-tuning \acp{LLM} for price prediction. For the third model we fine-tune a pre-trained RoBERTa-Base model \citep{liu_2019_roberta} directly on the cryptocurrency price. This model, in its raw form, isn't trained to perform any specific task yet, and needs to be fine-tuned. As the target for the training, we opt for daily price movements represented as a binary variable.

All three models handle emoticons (e.g.\ ``:)'') and unicode (e.g.\ emojis) appropriately and no additional vocabulary had to be added given that they already contain all relevant cryptocurrency-related vernacular in their pre-trained vocabulary. The cleaning of the textual data therefore only entails the removal of HTML elements and hyperlinks.

\begin{table}[!h]
	\centering
	\scriptsize
	\caption{Highest-scoring r/Bitcoin subreddit posts by NLP model}
	\renewcommand{\arraystretch}{1.1} 
	\begin{tabular}{p{0.21\textwidth}p{0.74\textwidth}}
		\toprule
		\textbf{Model} & \textbf{Highest-scoring r/Bitcoin post} \\
		\midrule
		VADER\vfill(sentiment dictionary) & Good $\vert$ Good \\
		\midrule
		Twitter-RoBERTa\vfill(sentiment model) & So excited I finally own 50 btc!! $\vert$ Thank you Bitcoin community! \\
		\midrule
		BART MNLI\vfill(bullishness model) & Foreign Exchange scandal will promote Bitcoin use! Getting screwed again by banks... $\vert$ German watchdog plans to step up FX probe at Deutsche. Britain's Financial Conduct Authority began a formal investigation into possible manipulation in the \$5.3 trillion-a-day global foreign exchange market. \\
		\midrule
		Fine-tuned RoBERTa \vfill(trained directly on the cryptocurrency price movements) & Ineligible to use the Coinbase platform $\vert$ I tried buying some coins just to hold on to, and I got an automated email saying my transaction was cancelled for security reasons. So I contacted support and they said: ``Unfortunately a manual review has determined that you are ineligible to use the Coinbase platform to purchase Bitcoin. We're sorry for any inconvenience that this may cause.'' Has this happened to anyone else? \\
		\bottomrule
	\end{tabular}
	\label{table:posts}
\end{table}

In Table \ref{table:posts}, we present the top-scoring r/Bitcoin post for the sentiment dictionary VADER and each of the three deep learning NLP models. The symbol `$\vert$' is used to separate the title of the Reddit post from the main body.

A post consisting merely of the word `Good' has achieved the highest rating according to the VADER sentiment dictionary. This is unsurprising, as `Good' is one of the highest-ranking words, and any additional words would only lower the average score, thereby highlighting the inherent limitations of dictionary-based approaches in sentiment analysis.

The Twitter-RoBERTa sentiment model effectively selected a notably positive post, while the BART MNLI bullishness classifier chose a post that conveys optimism with regard to the future price of Bitcoin. It is noteworthy that the latter does not explicitly include terms like `bullish' or `bull', demonstrating the model's capability to infer higher-level semantics from the presented hypothesis.

When evaluating the third deep learning model, it is important to recognise that predicting the daily price movements is a much more complex task than sentiment analysis. Therefore, the high-scoring posts of the fine-tuned RoBERTa \ac{LLM} do not necessarily convey positivity or optimism. The top post presented above suggests that perhaps the model has picked up on individuals expressing their intent to purchase the respective cryptocurrency.

\subsection{Choice of Target Variables and Trading Strategy}\label{sec:target_methods}

In our exploration of cryptocurrency price forecasting, we utilise the CryptoCompare price at midnight for the target creation. This choice is motivated by the robustness of their CCCAGG methodology, which averages prices from 301 cryptocurrency exchanges. The weighting of this average is influenced by both the 24 hour volume and the time elapsed since the last transaction, ensuring a comprehensive and timely representation of the market.

Our first forecasting target is the log price change for the subsequent day, treated as a continuous variable. The underlying trading strategy is straightforward: we buy the asset if the forecasted price change is positive and sell it if it is negative. This method offers the advantage of simplicity, but it also hinges on the precision of the continuous forecast.

Subsequently, we consider a binary representation of the next day's price change for our second target. Here, a price increase is coded as 1, while no change or a price decrease is represented as 0. The corresponding trading strategy is to buy if the prediction exceeds a certain threshold and to sell if it falls below.

For the local extrema analysis, we delve into an approach centred on local extrema, spanning observational intervals of $+$/$-$ 7 days, $+$/$-$ 14 days, or $+$/$-$ 21 days. We construct two binary variables that indicate whether a given timepoint is a local minimum or maximum within the set time interval. These variables become the target for two distinct binary classification models. The forecasts of these two models are then used to construct a trading strategy, which aims at purchasing the asset at the troughs and selling it at the peaks. In all three cases our trading simulation starts by purchasing the asset at the first timestep and ends with the liquidation of all held assets at the last timestep.

\begin{figure}[!h]
	\centering
	\hspace{-1.2em}\includegraphics[width=0.65\textwidth]{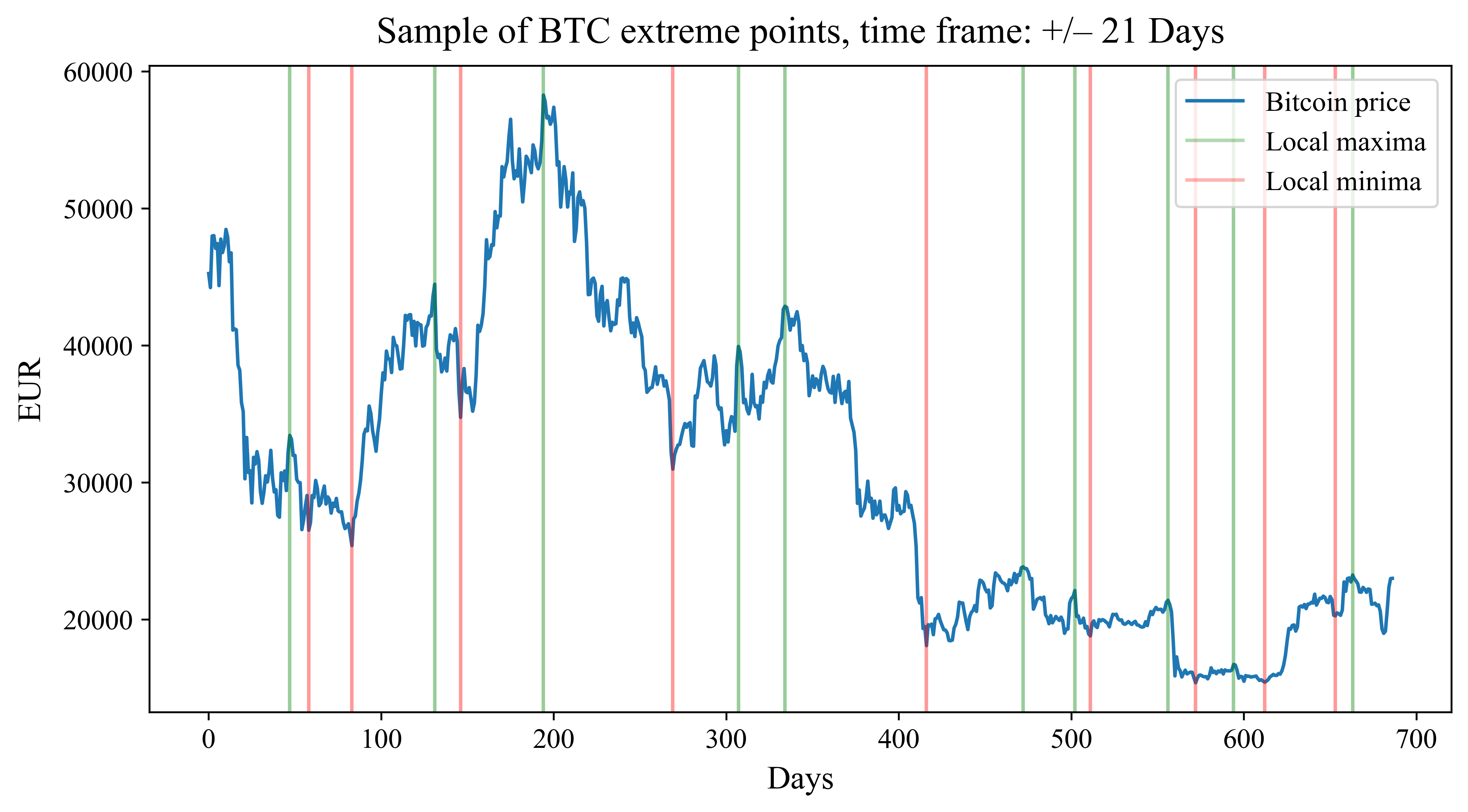}
	\caption{Sample of local extreme points of the \ac{BTC} price}
	\label{fig:extrema}
\end{figure}

By virtue of conveying less granular information compared to daily price fluctuations, local extrema as target variables imply a ceiling on potential profits. However, it is essential to reconceptualise this perspective. Given that local extrema are less susceptible to noise in comparison to daily price changes, an argument can be made that they present a more stable and distinct pattern for machine learning approaches to model. Thus, even though the extrema-based models might be associated with lower potential profit, their heightened accuracy could translate into greater profitability in practice. This contrasts with models trained on daily variations which, while encapsulating more information, might be impeded by their inherent noise and variability, leading to less efficient predictions that only generate a fraction of the potential profit. Furthermore, the prediction of local extreme points allows us to examine the impact of textual data across varying observational time frames, giving insight into the longevity of the effects of sentiments propagated through news and social media.

For the trading simulation, we buy if the predicted probability of a local minimum occurring the next day exceeds a set threshold, provided the predicted probability of a local maximum does not surpass the same threshold. Conversely, the strategy is to sell if the probability of a local maximum the next day exceeds the threshold, while the probability of a local minimum does not.

To optimise the efficacy of our strategies in the scope of the classification forecasts, we determine classification thresholds through a comprehensive grid search with the Accuracy metric as the objective. This metric proves to correlate well with maximum profit, while offering a significant computational advantage over the extensive resources required for running trading simulations at each threshold.


\section{Experimental Design}\label{sec:design}

\begin{figure}[!h]
	\centering
	\includegraphics[width=0.85\textwidth]{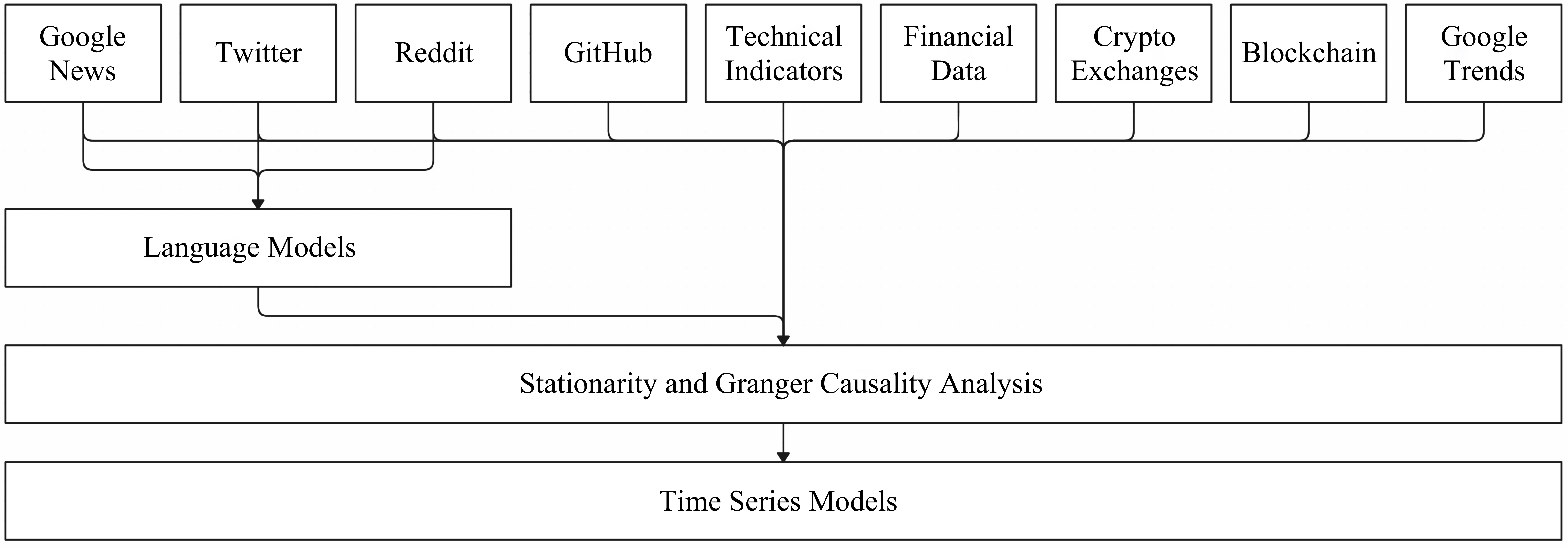}
	\vspace{0.6em}
	\caption{Overview of the experimental design}
	\label{fig:design}
\end{figure}

\subsection{Data Collection and Preprocessing}\label{sec:data}

We utilise a diverse range of data sources with the time frame of our dataset ranging from August 2011 for \ac{BTC} and from August 2015 for \ac{ETH} until March 2023. We collect text data from social media platforms and news outlets, focusing on English content. From Google News, we extract approximately 55 thousand news headlines, encompassing all articles from CoinDesk, Cointelegraph, and Decrypt that mention the keywords `Bitcoin' or `BTC' (and `Ethereum' or `ETH' respectively). On Reddit, we gather all posts from the r/Bitcoin and r/ethereum subreddits, totaling around 338 thousand threads. Finally, Twitter contributes the most to our dataset with nearly 1.9 million posts. We consider all tweets with more than five likes and two retweets that feature the hashtags \#bitcoin or \#btc (and correspondingly, \#ethereum or \#eth).

News headlines, with their formal and timely presentation of current affairs, offer a broad and credible overview of the latest events. Twitter, a popular platform among key influencers in the cryptocurrency domain, provides an unfiltered reflection of public opinion. With their concise format, tweets serve as a window into immediate personal reactions and insights, offering a snapshot of real-time sentiments. Meanwhile, Reddit posts, typically longer in nature, delve deeper into community-driven discussions, including background research and technical analysis. Together, these three sources provide a comprehensive blend of journalistic reporting, real-time reactions, and detailed community perspectives, making them invaluable for a holistic analysis of market sentiments and trends.

The text data is processed using the sentiment dictionary VADER and the three \acp{LLM} detailed in Section \ref{sec:nlp_methods}. In addition to the textual information, we consider numerical data from these platforms, such as the post count, the number of subscribers to the official Bitcoin/Ethereum Twitter accounts, and their respective subreddits, as well as the number of active users on Reddit.

Our study further integrates data from the code hosting platform GitHub, specifically the repositories of the two cryptocurrencies we analyse (`bitcoin/bitcoin' and `ethereum/go-ethereum'), offering a perspective on development activity. We consider the commit count, number of additions/deletions, forks, stars, and subscribers.

Considering financial data, we incorporate 48 different technical indicators based on past price and volume, including trend, momentum, volatility, and volume indicators. Additionally, we include the price and volume of the S\&P 500 index, the CBOE Volatility Index (VIX), COMEX gold price, and crypto indices such as the MarketVector Digital Assets 100 (MVDA) tracking the 100 largest cryptocurrencies and the Bitcoin Volatility Index (BVIN) that tracks the implied volatility of \ac{BTC} using options data from Deribit. Data from cryptocurrency exchanges, detailing the volume of purchases and sales of the cryptocurrencies of interest, is also factored into the analysis.

From the blockchain, we extract data on the amount and size of transactions, account balance data, the number of newly created addresses, the number of zero balance addresses, and other technical data such as the hashrate and block size.

Lastly, we leverage Google Trends to gauge public interest. We include the increase or decrease in the number of searches for the queries `bitcoin', `ethereum', `cryptocurrency', `blockchain', and `investing'. An extensive overview and description of all variables can be found in Table \ref{tab:variables} in Appendix B.

Values identified as erroneous due to technical anomalies, such as periods of social media platform downtime, are manually excluded. Apart from these specific instances, outliers are not removed from the dataset. The target variables and blockchain data are complete, containing no missing values. However, missing values are present in the financial data, notably prices on weekends, and are sporadically found in the social media and cryptocurrency exchange data due to server errors. These gaps are addressed by imputing the value from the previous day.

In terms of preprocessing, we first identify variables with a unit root using the Dickey-Fuller, Phillips-Perron, and Kwiatkowski-Phillips-Schmidt-Shin tests. For such variables, we take differences. Heteroskedastic variables, identified using the White, Breusch-Pagan, and Goldfeld-Quandt tests, are logged. We consider variables as non-unit root or homoskedastic if at least two of the three corresponding tests suggest so.

We consider lagged values of up to 14 days of all features, including financial indicators, and apply Granger causality for feature selection. Our decision to utilise 14 lags is underpinned by the fact that the presence of causal lags diminished considerably beyond the 14th day. To select features for our non-sequential models, we test for a Granger causal relationship between each lag of each feature with the respective target variable. The sequential models require all lags to be passed---thus, in this case, we test for Granger causality between all lags of each feature and the respective target.

Our Granger causality analysis reveals that the prices of both \ac{BTC} and \ac{ETH} are significantly influenced by their respective trading volumes, technical indicators, public sentiment, and broader economic trends. While both share common predictors, the \ac{BTC} price shows a more pronounced response to its network metrics and large transactions, whereas \ac{ETH} is affected more by developer activity. Examining the time frame of the impact of different features, both coins appear to be influenced by an interplay between real-time fluctuations and longer-term trends. Volume data and technical indicators have an immediate term impact of one or two days. Broader economic indicators, such as stock indices or the gold price, influence price movements over varying short-term intervals of one to two weeks. Meanwhile, \ac{NLP} data has a more diverse impact, from nearly immediate to up to a week later, demonstrating the varying half-lives of different text sources in influencing cryptocurrency prices.

Additionally, we perform seasonality decompositions of the differenced price data using MSTL \citep{bandara_2021_mstl} and Facebook's Prophet model \citep{taylor_2017_forecasting}. Both approaches reveal no significant daily, weekly, or monthly seasonality. Hence, we refrain from including day of the week or the month of the year dummies in the dataset. In total, we have at our disposal 137 variables, out of which between 52 and 84 are determined to be Granger causal, depending on the target.

\subsection{Model Development and Optimisation}\label{sec:models}

As the target for the fine-tuning of RoBERTa-Base, we opt for daily price movements represented as a binary variable. We strategically utilise the available textual data: half of the data from each day is allocated to the training process, while the remaining half is employed to compute the final scores. We fine-tune the hyperparameters of the RoBERTa model for each text source and each coin individually, given the significant differences in text length and stylistic characteristics. To this end, we employ the Bayesian optimisation framework Optuna \citep{akiba_2019_optuna}. The hyperparameter search entails 240 iterations with the \ac{AUC ROC} as the objective function. The search ranges of the hyperparameter tuning are outlined in Table \ref{tab:optuna_nlp} in Appendix B.

For the time series analysis, we employ a range of sequential and non-sequential forecasting models. We begin with OLS-based models for benchmarking, specifically Ridge Regression for the regression problems and Logistic Regression with L2 regularisation for the binary classification problems. Our findings indicate that L2 regularisation is superior to L1 in this context, offering a more robust model fit.

Given their capacity to model complex non-linear relationships, we also apply Gradient Boosting as implemented in the XGBoost framework and a vanilla \ac{MLP}. The objective functions for XGBoost are the \ac{MSE} for regression problems and \ac{BCE} for binary classification problems. Regularisation measures comprise a combination of L1 and L2 regularisation on the leaf weights, a threshold for the addition of new leaves to a tree (also referred to as `gamma'), as well as subsampling.

We construct the \ac{MLP} with a maximum of four layers and apply a parameter penalty using the L2 norm to mitigate overfitting. Unlike most approaches in the existing body of literature, we individually tune the number of neurons in each \ac{FNN} layer rather than setting a uniform count across all layers. This approach provides the models with an additional flexibility, optimising its adaptability to different data patterns. Aside from the neuron count, we tune as hyperparameters the activation function, the batch size, the learning rate, the optimiser, and the type of scaling.

Next, we construct an \ac{LSTM} architecture consisting of up to three \ac{LSTM} layers and an optional dense head with up to three dense layers. The hyperparameter tuning is very similar to that applied with the \ac{MLP}. Not only the \ac{LSTM} layer sizes are individually tuned, but also the neuron counts of the feed-forward layers that rest on top of them. The main difference in the tuning approach is the utilisation of `dropout', that is, the random deactivation of neurons during training, instead of L2 regularisation.

Finally, we explore the \ac{TFT}, a model which is particularly challenging due to its extensive training time. This is a result of it being fed all variables and conducting variable selection using \ac{GRN}s, a notably less efficient method than the Granger causality approach we adopt for all other models. Due to these time constraints, we opt for a uniform neuron count across all TFT layers. Aside from that, we employ dropout as the regularisation technique and set the number of attention heads relatively high, anticipating the complex seasonal patterns of our input time series. This assumption is validated as models with 16 attention heads consistently deliver superior performance.

For the sake of reproducibility, we abstain from employing early stopping in the training of the \ac{MLP}, \ac{LSTM}, and \ac{TFT} models. Instead, we treat the number of epochs as a tunable hyperparameter. For the regression task, we configure our neural-network-based models to use a linear activation in the output layer, with backpropagation driven by the \ac{MSE}. For the classification tasks, we employ a sigmoid activation in the output layer and use the \ac{BCE} loss for backpropagation. Given the inherently imbalanced nature of classifying local extrema, we apply reweighing to the minority class for all extreme point models.

For the hyperparameter tuning with Optuna, we use the trading profit as the objective function. The search ranges for this tuning are detailed in Table \ref{tab:optuna} in Appendix B. For each target and model, the optimisation process is either terminated after 200 iterations or after six weeks, whichever comes first. Notably, only the TFT ends up being constrained by the time limit.

\subsection{Performance Metrics and Model Evaluation}

We employ several evaluation metrics to assess the performance of our cryptocurrency forecasting models. Firstly, we utilise the \ac{AUC ROC} to measure the model's capability to rank positive instances higher than negative ones. Additionally, we measure the Accuracy of the model, which quantifies the proportion of correct predictions relative to the total number of predictions made. Beyond these traditional metrics, we introduce a practical evaluation based on the profitability of the model within the context of a trading strategy. To this end, we compare the profit and Sharpe ratio generated by our model-driven trading decisions to a buy-and-hold benchmark. This benchmark represents a passive investment strategy where an investor buys the asset and holds onto it for the entire duration of the time period. For the calculation of the Sharpe ratio we imply a risk-free rate of 0 \% and apply annualisation as described by \cite{sharpe_1994_the} for the sake of interpretability. 365 days are used for the annualisation since cryptocurrencies are also traded on weekends.

\begin{align*}
    &\textit{Annualised Sharpe Ratio} \,= \frac{365~\overline{r}}{\sqrt{365}~\sigma} \hspace{8pt} \text{ where } \hspace{9pt} \overline{r} = \frac{1}{n} \sum_{t=1}^n r_t ~\,, \hspace{8pt} \sigma = \sqrt{\frac{1}{n-1} \sum_{t=1}^n (r_t - \overline{r})^2}\\[5pt]
    &r_t := \text{ asset return on day } t\\[1pt]
    &n := \text{ total number of days in the given time window}
\end{align*}

For the profit calculation, we start with the assumption of a portfolio value of one euro. When our model anticipates a price increase or identifies a local minimum for the subsequent day, we invest the entire available capital to buy the asset. Conversely, if the model foresees a price drop or a local maximum the next day, we liquidate all held assets. The trading strategy does not involve short selling or investing in an alternative asset, after the cryptocurrency is sold. To further ensure simplicity and interpretability of our analysis, we do not account for transaction costs. This omission is justified by the emergence of off-chain systems, such as the Lightning and Raiden networks, which enable the trading of cryptocurrencies at significantly reduced transaction costs \citep{hafid_2020_scaling}.

\begin{figure}[!h]
    \centering
	\hspace{-0.7em}\includegraphics[width=0.55\textwidth]{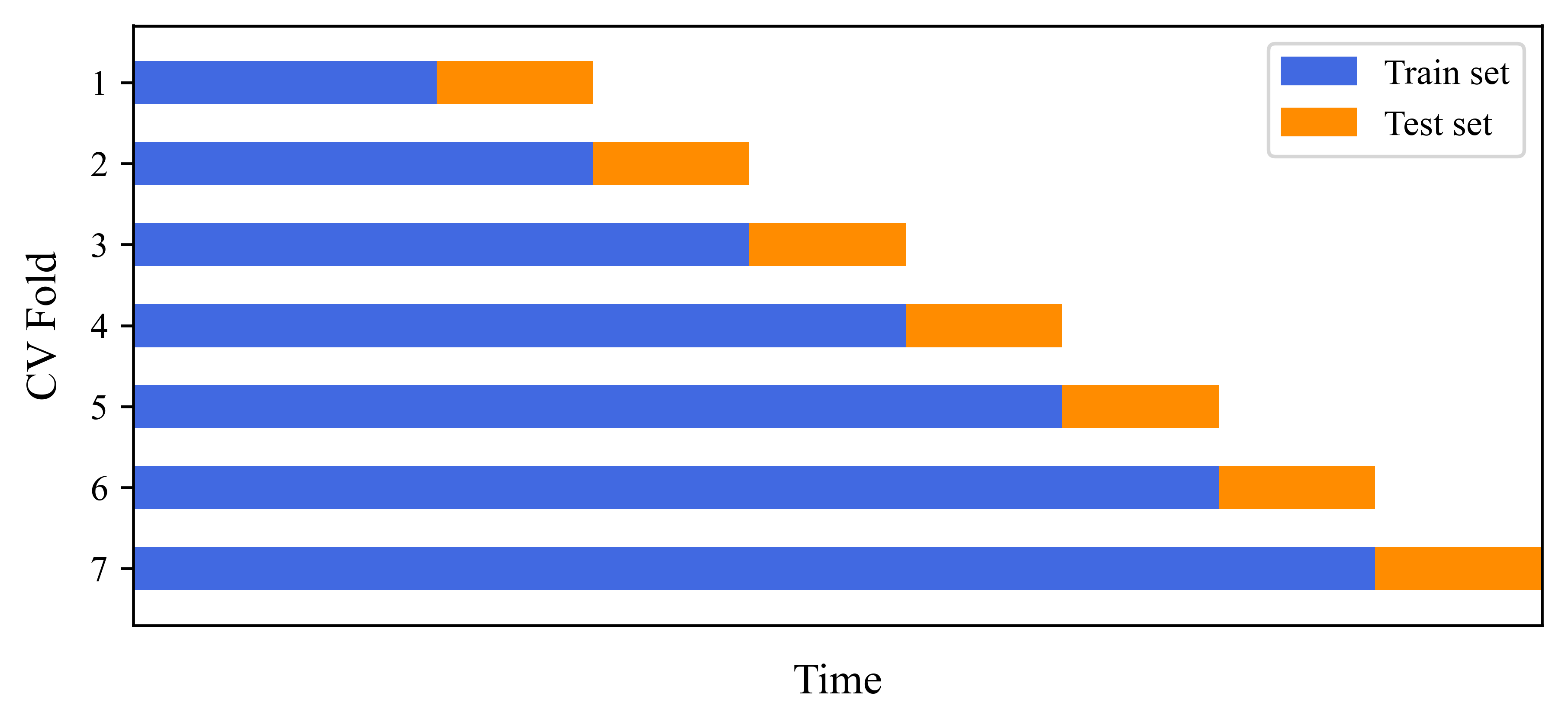}
	\caption{The applied time series cross-validation approach}
	\label{fig:cv}
\end{figure}

All the metrics are computed as averages of a 7-fold rolling-window cross-validation with incrementally increasing training window sizes (see Figure \ref{fig:cv}). The reasons for opting for increasing window sizes over a constant window, despite the higher computational demands, are twofold. Firstly, the increasing window approach is inherently more stable and results in lower variability of the computed metrics. Secondly, the models consistently exhibit superior performance when trained on the entirety of past data, as opposed to being limited to the most recent data points, suggesting that the underlying relationships have not changed significantly over time. The increasing window approach therefore provides a more accurate representation of model performance for the following comparative analysis.


\section{Results and Analysis}\label{sec:results}

\subsection{Comparison of Forecasting Performance}

In this section, we delve into a comprehensive examination of the BTC and ETH price forecasting performance. We apply a range of \ac{ML} models, described in detail in Section \ref{sec:models}, to five different target variables, explained in Section \ref{sec:target_methods}. Each model is trained once using financial, blockchain, GitHub, Google Trends, and numerical social media data, and then another time additionally incorporating various \ac{NLP} features. The subsequent analysis not only illuminates the potential profitability of different trading strategies but also evaluates the predictive power of NLP models in the context of financial forecasting.

In framing our subsequent analysis of trading profits, we start by considering a few reference points. Table \ref{tab:reference_points} outlines the profits resulting from implementing a buy-and-hold trading strategy and the profit resulting from trading given perfect knowledge of the respective target variable. All values are arithmetic means of our time series cross-validation approach, with each fold spanning a time frame of approximately 1.5 years, and were aggregated across both cryptocurrencies.

\begin{table}[!h]
	\scriptsize
	\centering
	\caption{Reference points for the analysis of trading profit}
	{\def\arraystretch{1.1}
		\begin{tabular}{p{0.24\textwidth}R{0.1\textwidth}R{0.1\textwidth}R{0.15\textwidth}R{0.11\textwidth}}
			\toprule
			Trading strategy
			& \multicolumn{2}{l}{\textbf{Buy-and-hold}}
			& \multicolumn{2}{l}{\textbf{Perfect knowledge of target}}
			\\
			\midrule
			& \multicolumn{1}{l}{Profit} & \multicolumn{1}{l}{\# of trades}
			& \multicolumn{1}{l}{Profit} & \multicolumn{1}{l}{\# of trades} \\
			\midrule
			Price movement (binary)    & 182.78 \%   & 2.0    & 99,092.42 \%  & 205.9     \\
			Extrema $+$/$-$ 7 days     & "           & "      & 759.40 \%     & 28.9      \\
			Extrema $+$/$-$ 14 days    & "           & "      & 761.64 \%     & 14.2      \\
			Extrema $+$/$-$ 21 days    & "           & "      & 702.27 \%     & 9.3       \\
			\bottomrule
		\end{tabular}
	}
	\label{tab:reference_points}
	\vspace{3pt}\newline
	{\fontsize{7pt}{7pt}\selectfont
		All metrics are averages of 7-fold cross-validation and were aggregated across both coins
	}
\end{table}

The buy-and-hold benchmark represents a passive investment strategy where the asset is bought and held for the entire duration of the respective cross-validation split. On the other hand, when guided by perfect knowledge of a target variable, a trader would purchase the asset ahead of every price surge and liquidate it prior to any decline. Such a strategy represents the upper bound for the potential profit of a target variable.

A striking observation is the immense profit potential linked to daily price movements, a characteristic rooted in the inherent volatility of cryptocurrency prices. Since a substantial proportion of these daily fluctuations can be attributed to random noise, it becomes crucial to evaluate how effectively our time series models can distill the information contained in these variables. It is interesting to assess whether the daily price movements emerge as the most profitable in practice or whether the extrema prove more insightful, despite their constrained profit ceiling.

By integrating the \ac{NLP} outputs as features into our time series models, we observe a clear improvement in forecasting performance. Not only does this integration substantially enhance the profitability, it also improves the \ac{AUC ROC} and the Accuracy.

\begin{figure}[!h]
	\includegraphics[width=0.95\linewidth]{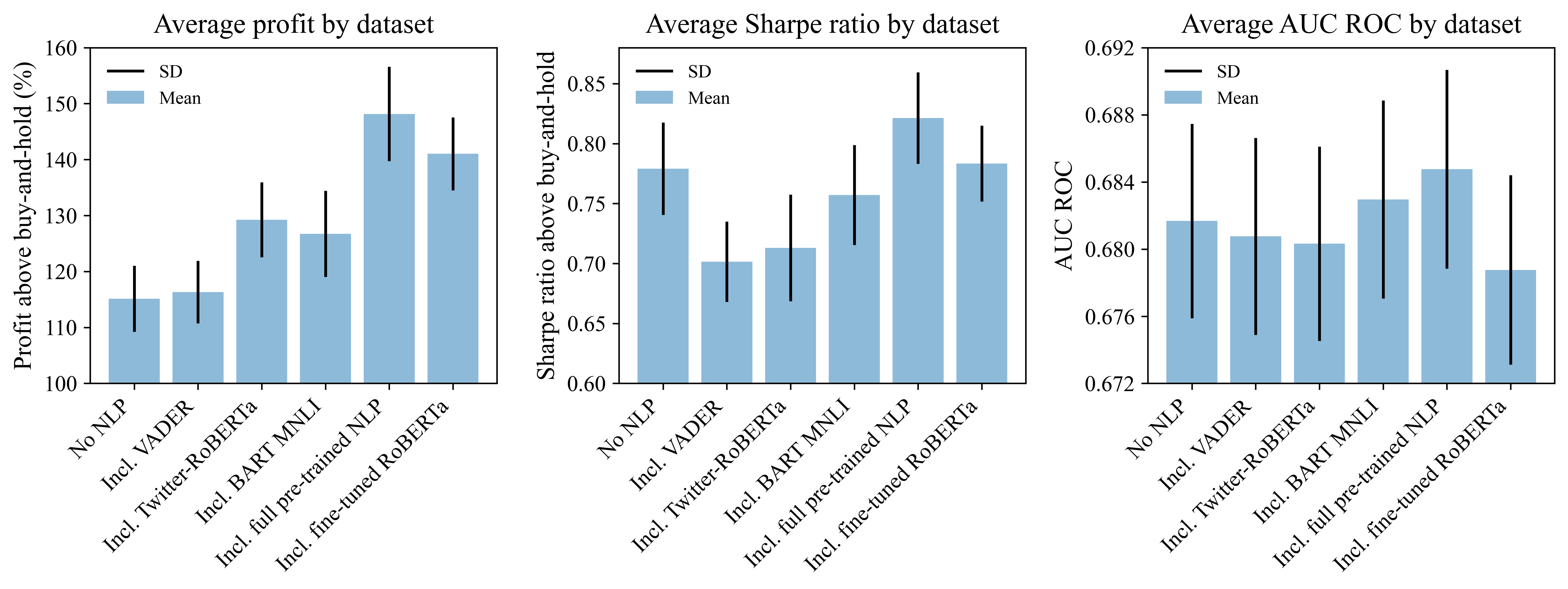}
	\caption{Comparison of \ac{MLP} profit, Sharpe ratio and AUC ROC by dataset}
	\label{fig:datasets}
\end{figure}

Fig. \ref{fig:datasets} offers a comparative analysis of the profit, Sharpe ratio and \ac{AUC ROC} of an \ac{MLP} model across the various sets of \ac{NLP} features, aggregated over both cryptocurrencies. In this context, ``full pre-trained NLP'' denotes the integration of the scores from both pre-trained \acp{LLM}: Twitter-RoBERTa and BART MNLI. The profit reported is the amount of percentage points above the profit resulting from a buy-and-hold strategy. For a more detailed examination of the significance of NLP data across the various cross-validation splits see Section \ref{sec:efficiency}.

The comparison vividly illustrates that deep learning \ac{NLP} models surpass the sentiment dictionary VADER, highlighting the advanced capabilities of these models. Despite its simplicity, VADER still contributes positively to forecasting profit. The Sharpe ratio, however, significantly diminishes when compared to using no NLP model, indicating a high level of noise in the VADER scores.

An intriguing observation is the interplay between the \ac{NLP} models used. While the Twitter-RoBERTa sentiment model and the BART MNLI bullishness classifier perform on par individually, integrating both models yields the highest performance in terms of all metrics. This confirms that our NLP models indeed extract different signals and indicates that previous studies have not fully exploited the informative value of text data for forecasting.

Another observation that deserves a special mention relates to the performance comparison between pre-trained and fine-tuned models. The pre-trained \ac{NLP} models, despite not being tailored to our dataset, yield greater benefits than those of the fine-tuned \ac{LLM}. Especially the comparatively low Sharpe ratio and AUC ROC indicate that the fine-tuned RoBERTa model introduced as much noise as it introduced information. This shows how complex the relationship between social media data and price fluctuations is and underscores the potential of transfer learning in the domain of financial forecasting.

\begin{figure}[!h]
	\includegraphics[width=0.95\linewidth]{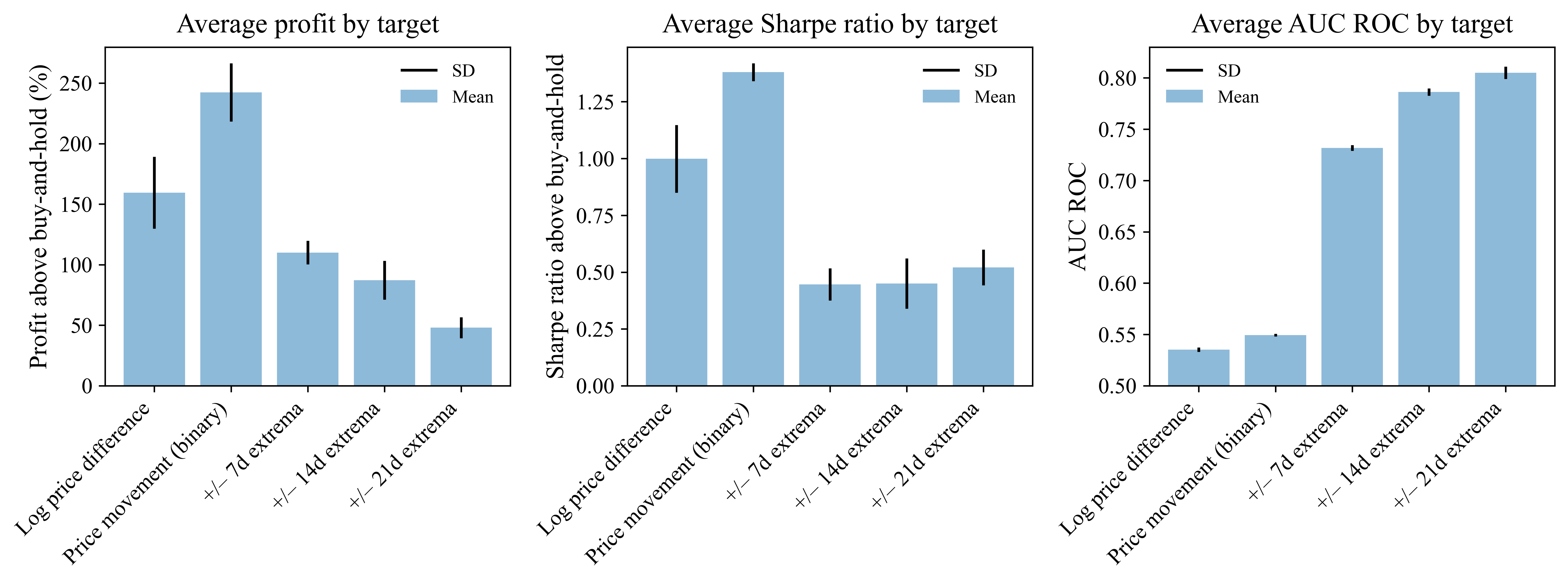}
	\caption{Comparison of \ac{MLP} profit, Sharpe ratio and AUC ROC by target variable}
	\label{fig:targets}
\end{figure}

Next, we examine the comparative performance of forecasting different targets. Fig. \ref{fig:targets} provides a visual representation, illustrating the average performance of an \ac{MLP} model broken down by target variables. A fundamental observation from our analysis is that all the chosen target types consistently outperform the buy-and-hold profit and Sharpe ratio accompanied by a decent \ac{AUC ROC}, thus underscoring their viability for financial forecasting.

However, we discover certain differences. The binary representation of the daily price change, which simplifies the price movements into two categories of increase or decrease, consistently surpasses its continuous counterpart. This result might be a consequence of the binary representation being less affected by market noise. We also observe, that under our assumption of zero frictions, the daily models demonstrate superior profitability compared to the extrema models, despite exhibiting higher levels of noise. This can be attributed to the large potential profit associated with them, as showcased in Table \ref{tab:reference_points}. Therefore, even though we are able to harness only a small fraction of the daily fluctuations, this fraction still contains more information than what we are able to extract from the local extreme points. It's important to note however, that the inclusion of transaction costs of merely 0.5 \% per transaction already positions the extrema models as equally competitive in terms of profitability, due to their significantly lower trading frequency. In the face of high transaction costs, the extrema approach might therefore not only be competitive, but even superior to forecasting daily price movements.

Shifting our focus to the \ac{AUC ROC}, the extreme point models show greater values compared with the daily price change models indicating significantly better discriminatory power. Additionally, when one puts the profits generated by the extreme point models into the context of the potential profits detailed in Table \ref{tab:reference_points}, it becomes apparent that they capture a greater proportion of the information that the targets contain.

It's also worth highlighting that the Sharpe ratio looks more favourable for the extrema models than the profit, particularly the 21-day window. The fact that the Sharpe ratio increases with the increasing observation time frames, while the profit decreases, proves that their more stable signal and the lower trading frequency do reduce the portfolio variance. This positions extreme points an efficient alternative to targets with higher granularity---not only in the face of high transaction costs, but also for volatile assets or turbulent market phases.

\begin{figure}[!h]
    \centering
    \begin{minipage}[b]{0.39\linewidth}
        \includegraphics[width=\linewidth]{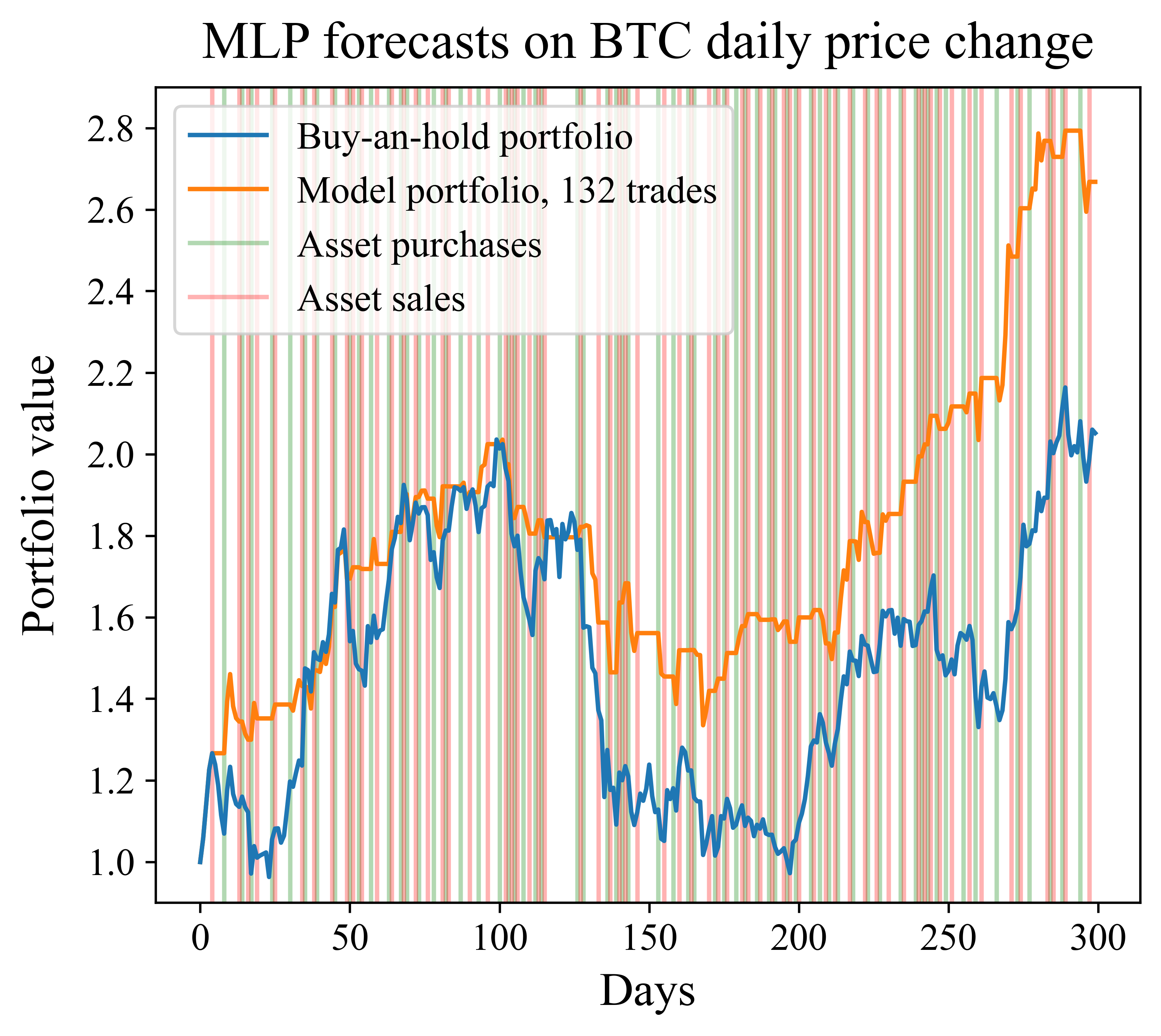}
    \end{minipage}
	\hspace{-0.01\linewidth}
    \begin{minipage}[b]{0.39\linewidth}
        \includegraphics[width=\linewidth]{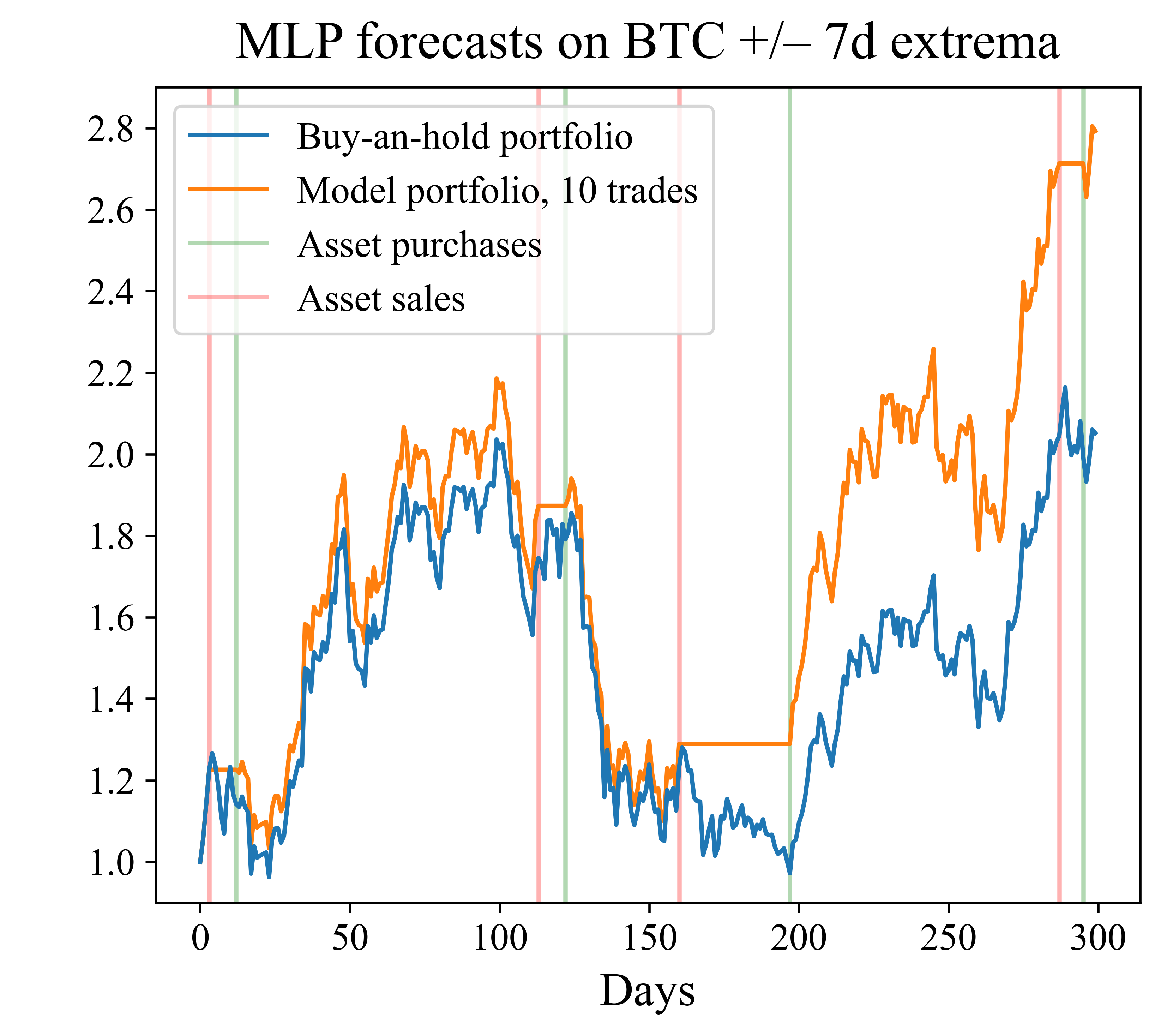}
    \end{minipage}
	\caption{Comparison of \acp{MLP} trained on daily price change and $+$/$-$ 7 day extreme points}
	\label{fig:sample_performance}
\end{figure}

In Fig. \ref{fig:sample_performance}, we illustrate the contrast between the two approaches by presenting a sample of the trading performances of \ac{MLP} classifiers, specifically one trained on daily price change in comparison to a prediction ensemble for $+$/$-$ 7 day extreme points. The blue line is the value of the asset and the orange line is the value of the trading portfolio, while the vertical lines indicate points of entry and exit. It's evident, that the extreme point models generate commendable results with significantly fewer trades than the model based on daily signals.

Expanding on our observations related to extrema prediction models, it becomes evident that not all market extremes carry equal weight. A few key turning points can have a bigger impact on profits than many smaller ones. If a model can accurately pinpoint these critical moments and skip the minor fluctuations, it stands to capture larger price movements, translating to more substantial profits. Our extreme point models seem to be adopting a ``quality over quantity'' approach, since it is those with fewer positive predictions (indicating extremes) and subsequently fewer trades that are the most profitable. For a detailed look at our findings, a thorough breakdown of the results can be found in tables \ref{tab:best_btc_reg} to \ref{tab:best_eth_21d} in Appendix A.

\begin{table}[!h]
	\centering
	\scriptsize
	\caption{Average performance by time series model}
	\begin{tabular}{p{0.14\textwidth}R{0.18\textwidth}R{0.12\textwidth}R{0.17\textwidth}R{0.14\textwidth}}
		\toprule
		\textbf{Model} & \textbf{Excess profit*} & \textbf{Trades**} & \textbf{AUC ROC**} & \textbf{Accuracy**} \\
		\midrule
		OLS/Logit & 67.48 \% & 52.7 & 0.6782 & 0.8025 \\
		XGBoost & 126.12 \% & 44.0 & 0.6998 & 0.8057 \\
		MLP (FNN) & 138.61 \% & 47.4 & 0.6797 & 0.8065 \\
		LSTM & 83.88 \% & 12.0 & 0.6526 & 0.8028 \\
		TFT & 11.13 \% & 4.2 & 0.5653 & 0.7971 \\
		\bottomrule
	\end{tabular}
	\label{tab:model_summary}
  \vspace{2pt}\newline
  {\fontsize{7pt}{7pt}\selectfont
  * Profit exceeding buy-and-hold strategy\\
  ** All metrics are averages of 7-fold cross-validation and were aggregated across all target variables and coins
  }
\end{table}

Navigating through the spectrum of models, we find that the OLS-based approaches already generate substantial and consistent profits as well as a decent \ac{AUC ROC}, possibly due to the complex non-linearities already captured by the financial indicators. Nevertheless, all models except for the TFT consistently surpass the OLS benchmark, with the MLP taking the lead. Regardless of the target variable in play, the MLP consistently emerges as the most profitable, simultaneously clocking the highest Accuracy. The XGBoost model on the other hand produces the highest \ac{AUC ROC}. Furthermore, it is the most profitable model for daily price movement prediction. A caveat worth noting here is that our hyperparameter tuning is aligned with profit optimisation. Thus, while the XGBoost model shines in terms of AUC ROC, the MLP might have outperformed XGBoost in that regard too, had it been specifically tuned with an emphasis on this metric.

Although the LSTM's overall performance trails slightly behind the MLP, it achieves its results with significantly fewer trades, adding a layer of efficiency. In particular, the LSTM displays good capability at forecasting extreme points. Given the inherent sequential characteristics of the LSTM, it requires the input of all lags for each selected variable though. Instead of performing Granger causality analysis for every individual lag, the procedure has to be streamlined by executing it for all lags of each variable simultaneously. Consequently, this approach supplies the LSTM with a higher volume of non-causal datapoints compared to the non-sequential models, which might be the cause for the LSTM's dampened performance.

In contrast, the weak performance of the TFT can likely be attributed to the fact that it utilises all lags of all variables and performs variable selection itself. This is significantly less efficient and, apparently, also less effective than the Granger causality approach employed for the other models, particularly given the large number of explanatory variables at hand. In addition, the reduced efficiency results in much greater training time and therefore unfortunately a less extensive hyperparameter tuning---again impacting performance. An emerging stream of literature, such as \cite{chen_2023_tsmixer}, has begun to critically evaluate the adoption of highly complex Transformer models for time series forecasting. They suggest that while Transformer architectures are powerful, simpler models, with better efficiency and interpretability, can often achieve comparable or even superior performance in time series forecasting, challenging the necessity of overly complex architectures.

\subsection{Analysis of Feature Importance}

To understand the significance of individual variables, we employ an XGBoost model trained on the entire set of available features. As prediction target, we utilise daily price movements encoded as a binary variable. This approach is justified by the performance of this model configuration, which achieved the highest \ac{AUC ROC} for both \ac{BTC} and \ac{ETH}, and was also ranked second and third in terms of profitability for each cryptocurrency, respectively.

We report average and total gain since they quantify the contribution a feature brings to the model's predictive capability. The average gain represents how beneficial, on average, splits on a specific feature are when they are made. Total gain, on the other hand, aggregates these benefits across all trees, representing a features' cumulative contribution to the model performance. For the sake of clarity and interpretability, we present the normalised values of these metrics. Portraying them as fractions of the total makes them interpretable as percentages of overall importance.

In Table \ref{tab:btc_importance} and \ref{tab:eth_importance}, the feature importances are aggregated for all lags of each feature. Moreover, given the many variables, we present the importance scores consolidated by feature category. Interested readers find disaggregated importance scores for individual features in Table \ref{tab:btc_importances} and Table \ref{tab:eth_importances} in the Online Resource.

\begin{table}[!h]
    \centering
	\scriptsize
    \caption{Feature categories ranked by their importance for predicting \ac{BTC} price movements}
	\begin{tabular}{p{0.3\textwidth}R{0.11\textwidth}R{0.11\textwidth}}
        \toprule
        \textbf{Feature category*} & \textbf{Normalised\vfill total gain} & \textbf{Normalised average gain} \\
        \midrule
		Technical indicators & 0.3722 & 0.4998 \\
		Transaction / account balance data & 0.2462 & 0.1558 \\
		NLP data & 0.1273 & 0.1224 \\
		Exchange volume data & 0.0753 & 0.0694 \\
		Technical blockchain metrics & 0.0556 & 0.0545 \\
		Numerical social media data & 0.0521 & 0.0366 \\
		Google Trends & 0.0262 & 0.0185 \\
		Past price data & 0.0235 & 0.0187 \\
		Financial data & 0.0168 & 0.0198 \\
		GitHub metrics & 0.0049 & 0.0046 \\
        \bottomrule
    \end{tabular}
	\label{tab:btc_importance}
  \vspace{2pt}\newline
  {\fontsize{7pt}{7pt}\selectfont* Categories are sorted by total gain}
\end{table}

Upon evaluating the \ac{BTC} feature importances, it is evident that technical indicators hold a preeminent position. The prominence of transaction and account balance data, ranked as the second most relevant category, highlights the valuable insights drawn from the transparent nature of individual wallet holdings. Additionally, the \ac{NLP} scores and post counts of Reddit and Twitter are noteworthy, emphasising the importance of textual data in financial forecasting. In particular, our fine-tuned RoBERTa model that was trained on the corpus of Tweets stands out by claiming the top position (as seen in Table \ref{tab:btc_importances} in Appendix A).

Technical metrics related to the blockchain, such as the hashrate or block size, also prove influential. Interestingly, of all the Google search query trends, ``blockchain'' ranks notably higher than the rest. This may suggest that a curiosity about blockchain's mechanics indicates a more profound interest in Bitcoin than just googling its name and, therefore, a higher likelihood of future purchase.

Other relevant features include past lags of price and volume as well as the circulating amount of the currency. Lastly, GitHub metrics appear to have the least impact on the model, suggesting that while the health of the developmental community in open-source projects is crucial, its bearing on the \ac{BTC} price might be minimal.

\begin{table}[!h]
    \centering
    \scriptsize
    \caption{Feature categories ranked by their importance for predicting \ac{ETH} price movements}
    \begin{tabular}{p{0.3\textwidth}R{0.11\textwidth}R{0.11\textwidth}}
        \toprule
        \textbf{Feature category*} & \textbf{Normalised\vfill total gain} & \textbf{Normalised average gain} \\
        \midrule
        Technical indicators & 0.5691 & 0.6714 \\ 
        NLP data & 0.2220 & 0.1657 \\ 
        GitHub metrics & 0.0541 & 0.0526 \\ 
        Exchange volume data & 0.0477 & 0.0334 \\ 
        Numerical social media data & 0.0341 & 0.0176 \\ 
        Transaction / account balance data & 0.0301 & 0.0207 \\ 
        Past price data & 0.0129 & 0.0097 \\ 
        Financial data & 0.0120 & 0.0121 \\ 
        Google Trends & 0.0091 & 0.0070 \\ 
        Technical blockchain metrics & 0.0089 & 0.0099 \\ 
        \bottomrule
    \end{tabular}
	\label{tab:eth_importance}
  \vspace{2pt}\newline
  {\fontsize{7pt}{7pt}\selectfont* Categories are sorted by total gain}
\end{table}

Examining the \ac{ETH} feature set, technical indicators persist in their dominance. Moreover, the significance of NLP models, particularly Twitter-RoBERTa and our fine-tuned RoBERTa model, is even more accentuated, reaffirming the overarching influence of social media on Ethereum's price dynamics. Other notable variables include the active user count on Reddit, trading data from various exchanges, and intriguingly, several metrics from GitHub, specifically the number of created and resolved issues, and the commit count. Being an indicator of upcoming technical changes, developmental activity may be of particular importance in the case of \ac{ETH}, considering its transition from a proof-of-work to a proof-of-stake consensus mechanism in 2022. Further variables of interest encompass transaction and account balance data as well as numerical social media data, such as the subscriber counts of the \ac{ETH} Twitter account or subreddit.

The prominence of technical indicators for both cryptocurrencies can be attributed to a number of factors. Firstly, while we provide the model with lags up to 14 of price and volume, some indicators can access a longer lookback period, thus encompassing more long-term information. Secondly, these indicators simplify intricate relationships into more digestible signals, making it easier for the model to discern patterns and trends that might otherwise be obscured in the raw data, especially considering our dataset's relatively limited size of a couple thousand observations. Thirdly, while the model only has access to the lags selected as relevant by Granger causality analysis, indicators like moving averages can combine the information of several consecutive lags, that may be missing in the feature set, in condensed form. Another dimension worth considering is the historical reliance of human traders on these indicators. If a significant section of market participants leans on these tools to make decisions, then the price movement will inherently reflect the signals from these indicators. Finally, the intrinsic smoothing within some of the used indicators can combat the noise in the raw data, acting as a form of implicit regularisation.

While our feature importance analysis underscores the significance of technical indicators, the outputs of our \ac{NLP} models, especially those representing Twitter and Reddit content, manifested as some of the most impactful explanatory variables for both \ac{BTC} and \ac{ETH}. This reaffirms our earlier conclusions that social media plays a pivotal role in influencing cryptocurrency price dynamics. Additionally, various data from the blockchain, exchange trading volumes, and metrics representing developer activity on GitHub have emerged as relevant.

\subsection{Market Efficiency Throughout Time}\label{sec:efficiency}

Market efficiency refers to the idea that asset prices in financial markets reflect all available information at any given time \citep{fama_1970_efficient}. In a market environment that is efficient, particularly in the semi-strong or strong form, consistently outperforming the market becomes challenging. This rapid incorporation offers minimal opportunities for traders to exploit the information for their advantage \citep[for an extensive review of the theoretical and empirical backgrounds of market efficiency see][]{shleifer_2009_inefficient}.

If a trader consistently achieves above-market profits, it could signify one of several situations: (i) the market is not efficient, (ii) the trader possesses a unique skill or system that the broader market has not adopted yet, or (iii) the trader is taking on higher risks to achieve those returns. Given that the profit of our trading portfolio is likely influenced by the latter two factors, we direct our attention to the trajectory of the profit over time, sidestepping the question whether the cryptocurrency market is efficient or not.

The time series models are evaluated by computing metrics on a 7-fold increasing window time-series cross-validation. By evaluating the model profits over the course of these seven cross-validation splits, we provide insights relating to the development of market efficiency throughout time.

Our evaluation will illuminate the consistency of the profits, hence providing an insight into the risk profile of the models' trading portfolio compared with the underlying cryptocurrency. Additionally, by observing if profits exhibit a trend over time, we can gauge if the market is increasingly integrating \ac{NLP} into their trading strategies, which might consequently shrink the potential future gains from text analysis.

\begin{figure}[!h]
    \centering
	\includegraphics[width=0.77\textwidth]{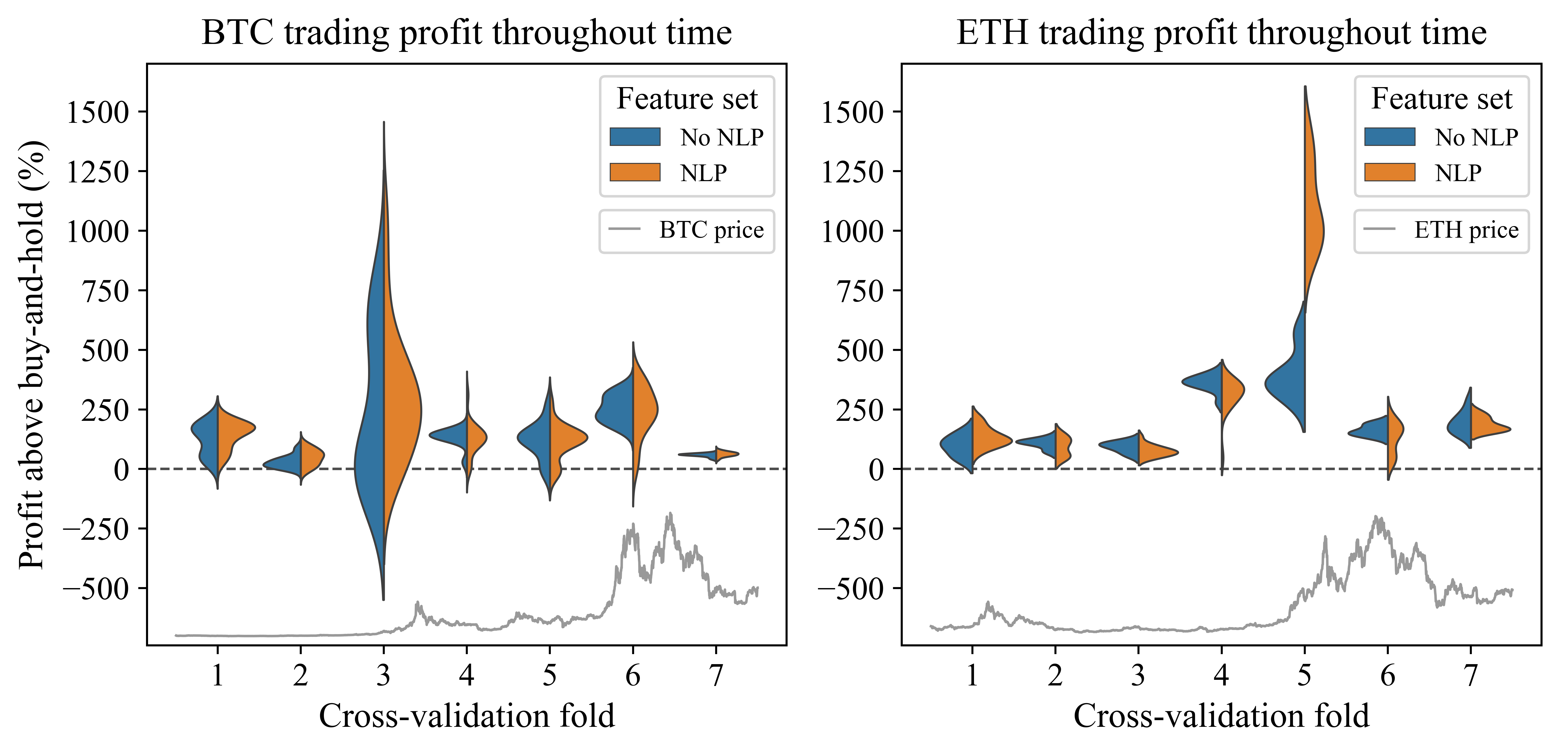}
	\caption{Distribution of trading profits of the 10 most profitable \ac{MLP} models throughout time}
	\label{fig:profit_by_split}
\end{figure}

Fig. \ref{fig:profit_by_split} displays the kernel density estimations for the trading profits of the 10 most profitable \ac{MLP} models across the cross-validation splits. The choice of focusing on the \ac{MLP} model was motivated by it being the most profitable among the \ac{ML} models.

We observe that our models consistently deliver trading profits that surpass the buy-and-hold benchmark. However, it is essential to highlight that while the profits remain largely above the benchmark for every cross-validation fold, the magnitude of profit does experience substantial fluctuations across the time splits. It is evident that during phases marked by heightened volatility of the underlying cryptocurrency, our models display a pronounced outperformance of the buy-and-hold approach.

We cannot observe a clear upwards or downwards trend in profits, whether in terms of \ac{NLP} effects or overall excess profit. This suggests that over the periods examined, the market's efficiency, or lack thereof, appears to remain largely unchanged. These findings indicate that textual data, as analysed through our methods, may not have significantly impacted market efficiency during the study period.

However, the violin plot sheds light on the significance of the impact of NLP data on our models' forecasting performance. For BTC, introducing NLP data into our models slightly nudges the profit distributions upwards for most splits, suggesting that the numerical representations derived from the textual data consistently provide information that is useful for predicting price movements. A second aspect is the potential reduction in volatility, most evident during the third split. This might indicate that linguistic data introduces more nuanced information, especially beneficial during turbulent market phases.

In the case of ETH, the benefits of NLP appear more period-specific, with notable advantages emerging in the fifth split, which was also characterised by the highest excess profit of the models trained without NLP data. Yet, in other splits, the NLP data seems less consequential, either offering limited enhancement or even marginally impeding the forecast. It is possible that during this specific period there was social media activity that was especially indicative of ETH's price movement. However, it seems more plausible to attribute the selective impact of the NLP data to the fact, that the fifth split is notably volatile and bullish. Given that these conditions present an elevated opportunity for ML models to leverage price fluctuations, they might have naturally become the primary target for our models, which were tuned with the objective of maximising trading profit.


\section{Conclusion}\label{sec:conclusion}

\subsection{Summary of Findings}

In this study, we explore the viability of news and social media data for cryptocurrency price forecasting. We are particularly interested in the time frame of the impact of textual data and the differences between various types of target variables. With regard to the targets for training, we utilise local extrema (minima and maxima) with varying observational time frames in addition to daily price movements. In the context of \ac{NLP}, we focus on investigating the application of multiple deep learning techniques. Moreover, we seek to assess the evolution of the market efficiency over time.

Our research reveals that including \ac{NLP} data improves the performance of our \ac{ML} models with respect to all evaluated metrics. Furthermore, deep learning \ac{NLP} models demonstrate superior performance compared with dictionary-based sentiment analysis. We find that pre-trained \acp{LLM}, namely Twitter-RoBERTa and BART MNLI, show promising capabilities in capturing market sentiment, performing on par with language models that are fine-tuned directly on the target at hand.

Additionally, our results indicate that text features lagged by up to one week are Granger causal and that incorporating \ac{NLP} data in the time series models results in enhancements in the forecasting of 21-day extrema. These findings suggest that news and social media can have a more long-term impact on price movements.

In terms of model performance, we find that non-linear models outperform those based on \ac{OLS}, demonstrating the existence of relevant non-linear relationships in the time series. We further identify that using the daily price change as a binary target variable consistently results in superior profitability compared to other targets, at least under the assumption of no transaction costs. Nevertheless, our models more reliably predict local extrema than daily price fluctuations, and the extreme point models yield decent profits and Sharpe ratio with significantly fewer trades.

All models consistently generate profits throughout all cross-validation splits; we do not observe a decrease in overall profits or a reduction in the impact of \ac{NLP} data across time. This suggests potential for the continued use of text analysis to enhance financial forecasts.

\subsection{Implications}

The incorporation of \ac{NLP} data into our models significantly improves price forecasting performance, demonstrating the value of considering such data in predictive efforts. Our study also underscores the efficacy of deep learning-based language models in this context. Particularly, when using several approaches in tandem, these models demonstrate significantly better performance compared with dictionary-based sentiment analysis.

Our results indicate that pre-trained models deliver comparable, if not superior, results to fine-tuned models, even when tackling abstract tasks in a specialised domain like finance. In particular, we find BART MNLI to be highly proficient as a zero-shot classifier. It effectively interprets market sentiment expressed through text that extends beyond mere positivity or negativity and substantially enhances predictive accuracy. These findings suggest promising prospects for the use of transfer learning in \ac{NLP} and not only highlight the versatility and robustness of pre-trained language models but also point towards a cost and time-effective route for future endeavours in financial forecasting. 

When turning our attention to the target variables, daily price movements encoded as a binary target consistently yield the highest profit. However, our models are also able to capture valuable information when employing local extrema as target variables. This suggests that although daily price movement may maximise profit, the use of extrema as target variables potentially offers deeper insights into the underlying market dynamics and proves useful in circumstances where one aims to reduce the number of trades, for example, in the face of high transaction costs.


\section*{Competing Interests Statement}

All authors certify that they have no affiliations with or involvement in any organisation or entity with any financial interest or non-financial interest in the subject matter or materials discussed in this manuscript.

\clearpage


\appendix

\section{Disaggregated Results}

\renewcommand{\thetable}{A.\arabic{table}} 
\setcounter{table}{0} 

\begin{table}[H]
	\centering
	\footnotesize
	\caption{The ten most profitable Bitcoin price movement regression models}
	\vspace{-0.5em}
	\begin{tabular}{p{0.112\textwidth}p{0.287\textwidth}R{0.127\textwidth}R{0.09\textwidth}R{0.118\textwidth}R{0.11\textwidth}}
		\toprule
		\textbf{Model} & \textbf{NLP features} & \textbf{Excess profit*} & \textbf{Trades**}  & \textbf{AUC ROC**} & \textbf{Accuracy**}\\
		\midrule
		XGBoost & Fine-tuned RoBERTa & 351.34 \% & 70.9 & 0.5262 & 0.5458 \\
		XGBoost & Twitter-RoBERTa & 317.41 \% & 134.4 & 0.5202 & 0.5409 \\
		XGBoost & Twitter-RoBERTa + BART MNLI & 300.96 \% & 94.0 & 0.5285 & 0.5394 \\
		XGBoost & None & 216.35 \% & 186.9 & 0.5242 & 0.5360 \\
		XGBoost & BART MNLI & 201.68 \% & 90.9 & 0.5198 & 0.5357 \\
		MLP (FNN) & BART MNLI & 172.08 \% & 142.3 & 0.5213 & 0.5379 \\
		LSTM & Fine-tuned RoBERTa & 159.98 \% & 22.9 & 0.5202 & 0.5400 \\
		LSTM & VADER & 139.93 \% & 8.6 & 0.5278 & 0.5357 \\
		LSTM & Twitter-RoBERTa & 126.24 \% & 61.1 & 0.5267 & 0.5375 \\
		LSTM & Twitter-RoBERTa + BART MNLI & 107.76 \% & 12.6 & 0.5213 & 0.5403 \\
		\bottomrule
	\end{tabular}
	\\\vspace*{0.7em}
	\scriptsize{* Profit exceeding buy-and-hold strategy;~~ ** All metrics are averages of 7-fold cross-validation}
	\label{tab:best_btc_reg}
\end{table}

\begin{table}[H]
	\centering
	\footnotesize
	\caption{The ten most profitable Ethereum price movement regression models}
	\vspace{-0.5em}
	\begin{tabular}{p{0.112\textwidth}p{0.287\textwidth}R{0.127\textwidth}R{0.09\textwidth}R{0.118\textwidth}R{0.11\textwidth}}
		\toprule
		\textbf{Model} & \textbf{NLP features} & \textbf{Excess profit*} & \textbf{Trades**}  & \textbf{AUC ROC**} & \textbf{Accuracy**}\\
		\midrule
		MLP (FNN) & Twitter-RoBERTa + BART MNLI & 251.98 \% & 127.4 & 0.5492 & 0.5529 \\
		MLP (FNN) & Fine-tuned RoBERTa & 230.31 \% & 104.9 & 0.5445 & 0.5481 \\
		MLP (FNN) & Twitter-RoBERTa & 224.38 \% & 109.4 & 0.5439 & 0.5495 \\
		MLP (FNN) & None & 210.75 \% & 128.3 & 0.5445 & 0.5462 \\
		MLP (FNN) & VADER & 208.85 \% & 80.9 & 0.5444 & 0.5481 \\
		XGBoost & Twitter-RoBERTa + BART MNLI & 210.53 \% & 108.3 & 0.5333 & 0.5333 \\
		MLP (FNN) & BART MNLI & 206.17 \% & 121.7 & 0.5466 & 0.5505 \\
		XGBoost & Twitter-RoBERTa & 200.77 \% & 112.9 & 0.5293 & 0.5300 \\
		XGBoost & Fine-tuned RoBERTa & 195.57 \% & 93.4 & 0.5254 & 0.5238 \\
		XGBoost & BART MNLI & 164.15 \% & 84.9 & 0.5314 & 0.5315 \\
		\bottomrule
	\end{tabular}
	\\\vspace*{0.7em}
	\scriptsize{* Profit exceeding buy-and-hold strategy;~~ ** All metrics are averages of 7-fold cross-validation}
	\label{tab:best_eth_reg}
\end{table}

\begin{table}[H]
	\centering
	\footnotesize
	\caption{The ten most profitable Bitcoin price movement classification models}
	\vspace{-0.5em}
	\begin{tabular}{p{0.112\textwidth}p{0.287\textwidth}R{0.127\textwidth}R{0.09\textwidth}R{0.118\textwidth}R{0.11\textwidth}}
		\toprule
		\textbf{Model} & \textbf{NLP features} & \textbf{Excess profit*} & \textbf{Trades**}  & \textbf{AUC ROC**} & \textbf{Accuracy**}\\
		\midrule
		LSTM & Twitter-RoBERTa & 234.35 \% & 13.7 & 0.5253 & 0.5525 \\
		XGBoost & BART MNLI & 217.23 \% & 116.0 & 0.5438 & 0.5656 \\
		MLP (FNN) & Twitter-RoBERTa & 214.29 \% & 92.3 & 0.5378 & 0.5620 \\
		MLP (FNN) & Twitter-RoBERTa + BART MNLI & 208.54 \% & 90.6 & 0.5339 & 0.5565 \\
		XGBoost & Twitter-RoBERTa & 203.69 \% & 143.7 & 0.5458 & 0.5696 \\
		XGBoost & Twitter-RoBERTa + BART MNLI & 202.01 \% & 157.7 & 0.5488 & 0.5659 \\
		LSTM & Fine-tuned RoBERTa & 193.66 \% & 17.1 & 0.5277 & 0.5519 \\
		XGBoost & Fine-tuned RoBERTa & 184.13 \% & 132.6 & 0.5408 & 0.5623 \\
		MLP (FNN) & None & 183.08 \% & 99.1 & 0.5354 & 0.5583 \\
		XGBoost & None & 177.05 \% & 61.7 & 0.5418 & 0.5653 \\
		\bottomrule
	\end{tabular}
	\\\vspace*{0.7em}
	\scriptsize{* Profit exceeding buy-and-hold strategy;~~ ** All metrics are averages of 7-fold cross-validation}
	\label{tab:best_btc_clf}
\end{table}

\begin{table}[H]
	\centering
	\footnotesize
	\caption{The ten most profitable Ethereum price movement classification models}
	\vspace{-0.5em}
	\begin{tabular}{p{0.112\textwidth}p{0.287\textwidth}R{0.127\textwidth}R{0.09\textwidth}R{0.118\textwidth}R{0.11\textwidth}}
		\toprule
		\textbf{Model} & \textbf{NLP features} & \textbf{Excess profit*} & \textbf{Trades**}  & \textbf{AUC ROC**} & \textbf{Accuracy**}\\
		\midrule
		XGBoost & Fine-tuned RoBERTa & 370.77 \% & 94.9 & 0.5607 & 0.5729 \\
		MLP (FNN) & BART MNLI & 354.81 \% & 82.9 & 0.5630 & 0.5705 \\
		MLP (FNN) & Twitter-RoBERTa + BART MNLI & 348.69 \% & 85.7 & 0.5612 & 0.5695 \\
		MLP (FNN) & Fine-tuned RoBERTa & 316.07 \% & 82.6 & 0.5660 & 0.5695 \\
		XGBoost & None & 311.69 \% & 103.4 & 0.5614 & 0.5705 \\
		Logit & BART MNLI & 304.98 \% & 77.1 & 0.5470 & 0.5676 \\
		Logit & Twitter-RoBERTa + BART MNLI & 302.06 \% & 80.0 & 0.5586 & 0.5710 \\
		Logit & Twitter-RoBERTa & 300.73 \% & 80.0 & 0.5650 & 0.5752 \\
		Logit & Fine-tuned RoBERTa & 295.88 \% & 67.7 & 0.5574 & 0.5690 \\
		XGBoost & Twitter-RoBERTa + BART MNLI & 291.43 \% & 98.6 & 0.5616 & 0.5714 \\
		\bottomrule
	\end{tabular}
	\\\vspace*{0.7em}
	\scriptsize{* Profit exceeding buy-and-hold strategy;~~ ** All metrics are averages of 7-fold cross-validation}
	\label{tab:best_eth_clf}
\end{table}

\begin{table}[H]
	\centering
	\footnotesize
	\caption{The ten most profitable Bitcoin +/-- 7 day extrema classification models}
	\vspace{-0.5em}
	\begin{tabular}{p{0.112\textwidth}p{0.287\textwidth}R{0.127\textwidth}R{0.09\textwidth}R{0.118\textwidth}R{0.11\textwidth}}
		\toprule
		\textbf{Model} & \textbf{NLP features} & \textbf{Excess profit*} & \textbf{Trades**}  & \textbf{AUC ROC**} & \textbf{Accuracy**}\\
		\midrule
		MLP (FNN) & Fine-tuned RoBERTa & 169.12 \% & 2.3 & 0.7334 & 0.9574 \\
		MLP (FNN) & BART MNLI & 140.25 \% & 2.6 & 0.7414 & 0.9577 \\
		MLP (FNN) & None & 135.93 \% & 2.6 & 0.7409 & 0.9576 \\
		MLP (FNN) & VADER & 134.45 \% & 2.6 & 0.7459 & 0.9577 \\
		MLP (FNN) & Twitter-RoBERTa + BART MNLI & 133.59 \% & 2.3 & 0.7396 & 0.9576 \\
		LSTM & Twitter-RoBERTa + BART MNLI & 129.85 \% & 2.3 & 0.6923 & 0.9570 \\
		LSTM & VADER & 128.22 \% & 2.3 & 0.6926 & 0.9568 \\
		MLP (FNN) & Twitter-RoBERTa & 127.43 \% & 2.3 & 0.7338 & 0.9574 \\
		LSTM & None & 126.26 \% & 2.0 & 0.6917 & 0.9570 \\
		LSTM & BART MNLI  & 126.26 \% & 2.0 & 0.6851 & 0.9570 \\
		\bottomrule
	\end{tabular}
	\\\vspace*{0.7em}
	\scriptsize{* Profit exceeding buy-and-hold strategy;~~ ** All metrics are averages of 7-fold cross-validation}
	\label{tab:best_btc_7d}
\end{table}

\begin{table}[H]
	\centering
	\footnotesize
	\caption{The ten most profitable Ethereum +/-- 7 day extrema classification models}
	\vspace{-0.5em}
	\begin{tabular}{p{0.112\textwidth}p{0.287\textwidth}R{0.127\textwidth}R{0.09\textwidth}R{0.118\textwidth}R{0.11\textwidth}}
		\toprule
		\textbf{Model} & \textbf{NLP features} & \textbf{Excess profit*} & \textbf{Trades**}  & \textbf{AUC ROC**} & \textbf{Accuracy**}\\
		\midrule
		LSTM & BART MNLI & 82.67 \% & 2.6 & 0.6890 & 0.9562 \\
		LSTM & None & 74.77 \% & 2.3 & 0.6814 & 0.9564 \\
		MLP (FNN) & Twitter-RoBERTa & 73.56 \% & 2.0 & 0.7221 & 0.9567 \\
		LSTM & Twitter-RoBERTa + BART MNLI & 73.52 \% & 2.0 & 0.6900 & 0.9564 \\
		MLP (FNN) & Twitter-RoBERTa + BART MNLI & 70.50 \% & 2.0 & 0.7270 & 0.9569 \\
		MLP (FNN) & Fine-tuned RoBERTa & 68.14 \% & 2.0 & 0.7230 & 0.9567 \\
		LSTM & Twitter-RoBERTa & 68.14 \% & 2.0 & 0.6779 & 0.9562 \\
		LSTM & Fine-tuned RoBERTa & 67.31 \% & 2.6 & 0.6737 & 0.9564 \\
		MLP (FNN) & BART MNLI & 49.65 \% & 2.0 & 0.7128 & 0.9564 \\
		MLP (FNN) & None & 49.41 \% & 2.0 & 0.7270 & 0.9567 \\
		\bottomrule
	\end{tabular}
	\\\vspace*{0.7em}
	\scriptsize{* Profit exceeding buy-and-hold strategy;~~ ** All metrics are averages of 7-fold cross-validation}
	\label{tab:best_eth_7d}
\end{table}

\begin{table}[H]
	\centering
	\footnotesize
	\caption{The ten most profitable Bitcoin +/-- 14 day extrema classification models}
	\vspace{-0.5em}
	\begin{tabular}{p{0.112\textwidth}p{0.287\textwidth}R{0.127\textwidth}R{0.09\textwidth}R{0.118\textwidth}R{0.11\textwidth}}
		\toprule
		\textbf{Model} & \textbf{NLP features} & \textbf{Excess profit*} & \textbf{Trades**}  & \textbf{AUC ROC**} & \textbf{Accuracy**}\\
		\midrule
		MLP (FNN) & Fine-tuned RoBERTa & 123.68 \% & 2.9 & 0.7581 & 0.9795 \\
		MLP (FNN) & VADER & 101.60 \% & 3.1 & 0.7632 & 0.9795 \\
		MLP (FNN) & Twitter-RoBERTa & 71.56 \% & 2.3 & 0.7628 & 0.9794 \\
		MLP (FNN) & None & 52.34 \% & 2.6 & 0.7616 & 0.9795 \\
		MLP (FNN) & Twitter-RoBERTa + BART MNLI & 46.92 \% & 2.6 & 0.7617 & 0.9795 \\
		XGBoost & Twitter-RoBERTa & 31.76 \% & 2.3 & 0.8069 & 0.9795 \\
		XGBoost & BART MNLI & 31.76 \% & 2.3 & 0.7986 & 0.9799 \\
		XGBoost & Fine-tuned RoBERTa & 27.28 \% & 2.3 & 0.8060 & 0.9795 \\
		XGBoost & None & 26.91 \% & 2.3 & 0.8031 & 0.9794 \\
		LSTM & VADER & 21.00 \% & 2.6 & 0.7159 & 0.9788 \\
		\bottomrule
	\end{tabular}
	\\\vspace*{0.7em}
	\scriptsize{* Profit exceeding buy-and-hold strategy;~~ ** All metrics are averages of 7-fold cross-validation}
	\label{tab:best_btc_14d}
\end{table}

\begin{table}[H]
	\centering
	\footnotesize
	\caption{The ten most profitable Ethereum +/-- 14 day extrema classification models}
	\vspace{-0.5em}
	\begin{tabular}{p{0.112\textwidth}p{0.287\textwidth}R{0.127\textwidth}R{0.09\textwidth}R{0.118\textwidth}R{0.11\textwidth}}
		\toprule
		\textbf{Model} & \textbf{NLP features} & \textbf{Excess profit*} & \textbf{Trades**}  & \textbf{AUC ROC**} & \textbf{Accuracy**}\\
		\midrule
		LSTM & BART MNLI & 105.21 \% & 2.3 & 0.7735 & 0.9810 \\
		MLP (FNN) & Twitter-RoBERTa + BART MNLI & 104.91 \% & 2.9 & 0.8144 & 0.9819 \\
		MLP (FNN) & Fine-tuned RoBERTa & 102.19 \% & 2.9 & 0.7998 & 0.9821 \\
		MLP (FNN) & BART MNLI & 101.97 \% & 2.3 & 0.7957 & 0.9819 \\
		LSTM & Twitter-RoBERTa & 95.87 \% & 2.3 & 0.7527 & 0.9814 \\
		MLP (FNN)  & Twitter-RoBERTa & 95.55 \% & 4.6 & 0.8159 & 0.9821 \\
		MLP (FNN)  & None & 69.31 \% & 2.6 & 0.8085 & 0.9826 \\
		XGBoost & Twitter-RoBERTa & 64.22 \% & 2.3 & 0.8325 & 0.9821 \\
		LSTM & VADER & 62.03 \% & 2.0 & 0.7581 & 0.9812 \\
		XGBoost & Twitter-RoBERTa + BART MNLI & 53.28 \% & 2.6 & 0.8334 & 0.9821 \\
		\bottomrule
	\end{tabular}
	\\\vspace*{0.7em}
	\scriptsize{* Profit exceeding buy-and-hold strategy;~~ ** All metrics are averages of 7-fold cross-validation}
	\label{tab:best_eth_14d}
\end{table}

\begin{table}[H]
	\centering
	\footnotesize
	\caption{The ten most profitable Bitcoin +/-- 21 day extrema classification models}
	\vspace{-0.5em}
	\begin{tabular}{p{0.112\textwidth}p{0.287\textwidth}R{0.127\textwidth}R{0.09\textwidth}R{0.118\textwidth}R{0.11\textwidth}}
		\toprule
		\textbf{Model} & \textbf{NLP features} & \textbf{Excess profit*} & \textbf{Trades**}  & \textbf{AUC ROC**} & \textbf{Accuracy**}\\
		\midrule
		LSTM & Twitter-RoBERTa + BART MNLI & 81.72 \% & 4.0 & 0.7565 & 0.9872 \\
		MLP (FNN) & VADER & 68.95 \% & 2.9 & 0.7939 & 0.9873 \\
		MLP (FNN) & BART MNLI & 47.08 \% & 2.3 & 0.8207 & 0.9875 \\
		MLP (FNN) & Fine-tuned RoBERTa & 36.24 \% & 2.3 & 0.7944 & 0.9875 \\
		MLP (FNN) & Twitter-RoBERTa + BART MNLI & 32.09 \% & 2.3 & 0.8012 & 0.9875 \\
		LSTM & VADER & 26.53 \% & 7.5 & 0.7572 & 0.9867 \\
		XGBoost & Twitter-RoBERTa + BART MNLI & 20.38 \% & 2.3 & 0.8487 & 0.9875 \\
		XGBoost & Fine-tuned RoBERTa & 19.14 \% & 2.3 & 0.8487 & 0.9875 \\
		XGBoost & Twitter-RoBERTa & 16.97 \% & 2.3 & 0.8517 & 0.9873 \\
		XGBoost & BART MNLI & 13.73 \% & 2.3 & 0.8567 & 0.9873 \\
		\bottomrule
	\end{tabular}
	\\\vspace*{0.7em}
	\scriptsize{* Profit exceeding buy-and-hold strategy;~~ ** All metrics are averages of 7-fold cross-validation}
	\label{tab:best_btc_21d}
\end{table}

\begin{table}[H]
	\centering
	\footnotesize
	\caption{The ten most profitable Ethereum +/-- 21 day extrema classification models}
	\vspace{-0.5em}
	\begin{tabular}{p{0.112\textwidth}p{0.287\textwidth}R{0.127\textwidth}R{0.09\textwidth}R{0.118\textwidth}R{0.11\textwidth}}
		\toprule
		\textbf{Model} & \textbf{NLP features} & \textbf{Excess profit*} & \textbf{Trades**}  & \textbf{AUC ROC**} & \textbf{Accuracy**}\\
		\midrule
		LSTM & Twitter-RoBERTa & 121.54 \% & 2.3 & 0.8260 & 0.9857 \\
		LSTM & VADER & 84.99 \% & 2.6 & 0.7730 & 0.9855 \\
		LSTM & Twitter-RoBERTa + BART MNLI & 79.57 \% & 2.3 & 0.8014 & 0.9860 \\
		LSTM & BART MNLI & 78.61 \% & 2.0 & 0.7971 & 0.9862 \\
		MLP (FNN) & Twitter-RoBERTa & 75.58 \% & 3.1 & 0.8088 & 0.9855 \\
		LSTM & Fine-tuned RoBERTa & 75.02 \% & 2.0 & 0.8162 & 0.9862 \\
		LSTM & None & 68.51 \% & 2.0 & 0.8123 & 0.9862 \\
		MLP (FNN) & BART MNLI & 55.70 \% & 2.6 & 0.7927 & 0.9862 \\
		MLP (FNN) & Twitter-RoBERTa + BART MNLI & 51.67 \% & 2.3 & 0.8057 & 0.9862 \\
		MLP (FNN) & None & 50.79 \% & 3.4 & 0.8133 & 0.9864 \\
		\bottomrule
	\end{tabular}
	\\\vspace*{0.7em}
	\scriptsize{* Profit exceeding buy-and-hold strategy;~~ ** All metrics are averages of 7-fold cross-validation}
	\label{tab:best_eth_21d}
\end{table}

\begin{table}[H]
    \centering
    \footnotesize
    \caption{The 50 most influential features for predicting BTC price movements using XGBoost}
    \begin{tabular}{p{0.4\textwidth} p{0.27\textwidth} R{0.09\textwidth} R{0.11\textwidth}}
        \toprule
        \textbf{Feature name*} & \textbf{Category} & \textbf{Normalised total gain} & \textbf{Normalised average gain} \\
        \midrule
        tweets\_roberta\_finetuned\_score & NLP data & 0.0754 & 0.0728 \\
        balance\_distribution\_from\_0.01\_addressesCount & Transaction / account balance data & 0.0430 & 0.0285 \\
        reddit\_count & Numerical social media data & 0.0346 & 0.0223 \\
        tweets\_twitter\_roberta\_pretrained\_score & NLP data & 0.0336 & 0.0343 \\
        indicator\_UI & Technical indicators & 0.0331 & 0.0278 \\
        average\_transaction\_value & Transaction / account balance data & 0.0308 & 0.0156 \\
        indicator\_AO & Technical indicators & 0.0291 & 0.0434 \\
        indicator\_Ichimoku\_A & Technical indicators & 0.0272 & 0.0294 \\
        balance\_distribution\_from\_0.01\_totalVolume & Transaction / account balance data & 0.0260 & 0.0181 \\
        price\_close & Past price data & 0.0235 & 0.0187 \\
        current\_supply & Technical blockchain metrics & 0.0204 & 0.0126 \\
        indicator\_NVI & Technical indicators & 0.0200 & 0.0338 \\
        total\_volume & Exchange volume data & 0.0198 & 0.0167 \\
        indicator\_Ichimoku\_Conversion & Technical indicators & 0.0197 & 0.0241 \\
        gtrends\_blockchain\_relative\_change & Google Trends & 0.0191 & 0.0134 \\
        indicator\_EMA & Technical indicators & 0.0189 & 0.0222 \\
        EUR\_volumefrom & Exchange volume data & 0.0181 & 0.0120 \\
        zero\_balance\_addresses\_all\_time & Transaction / account balance data & 0.0174 & 0.0099 \\
        indicator\_CR & Technical indicators & 0.0172 & 0.0148 \\
        balance\_distribution\_from\_100.0\_addressesCount & Transaction / account balance data & 0.0170 & 0.0101 \\
        indicator\_MACD & Technical indicators & 0.0165 & 0.0344 \\
        indicator\_DCM & Technical indicators & 0.0164 & 0.0217 \\
        indicator\_PPO & Technical indicators & 0.0160 & 0.0376 \\
        balance\_distribution\_from\_10.0\_addressesCount & Transaction / account balance data & 0.0154 & 0.0095 \\
        reddit\_bart\_mnli\_bullish\_score & NLP data & 0.0149 & 0.0093 \\
        indicator\_Vortex\_down & Technical indicators & 0.0148 & 0.0096 \\
        indicator\_KCW & Technical indicators & 0.0145 & 0.0088 \\
        unique\_addresses\_all\_time & Transaction / account balance data & 0.0135 & 0.0093 \\
        large\_transaction\_count & Transaction / account balance data & 0.0134 & 0.0092 \\
        indicator\_Stoch\_RSI & Technical indicators & 0.0131 & 0.0096 \\
        block\_time & Technical blockchain metrics & 0.0128 & 0.0098 \\
        indicator\_KCM & Technical indicators & 0.0126 & 0.0186 \\
        tweet\_count & Numerical social media data & 0.0120 & 0.0097 \\
        balance\_distribution\_from\_0.1\_totalVolume & Transaction / account balance data & 0.0119 & 0.0088 \\
        indicator\_BBM & Technical indicators & 0.0110 & 0.0167 \\
        hashrate & Technical blockchain metrics & 0.0110 & 0.0097 \\
        indicator\_KAMA & Technical indicators & 0.0108 & 0.0104 \\
        gold\_usd\_price & Financial data & 0.0093 & 0.0083 \\
        indicator\_BBW & Technical indicators & 0.0092 & 0.0049 \\
        indicator\_Ichimoku\_Base & Technical indicators & 0.0090 & 0.0171 \\
        block\_size & Technical blockchain metrics & 0.0089 & 0.0047 \\
        balance\_distribution\_from\_0.0\_addressesCount & Transaction / account balance data & 0.0087 & 0.0046 \\
        EUR\_volumeto & Exchange volume data & 0.0086 & 0.0091 \\
        indicator\_TRIX & Technical indicators & 0.0085 & 0.0178 \\
        balance\_distribution\_from\_100.0\_totalVolume & Transaction / account balance data & 0.0076 & 0.0043 \\
        indicator\_WMA & Technical indicators & 0.0076 & 0.0090 \\
        balance\_distribution\_from\_1000.0\_totalVolume & Transaction / account balance data & 0.0076 & 0.0051 \\
        indicator\_DPO & Technical indicators & 0.0075 & 0.0047 \\
        balance\_distribution\_from\_10.0\_totalVolume & Transaction / account balance data & 0.0072 & 0.0044 \\
        balance\_distribution\_from\_1000.0\_addressesCount & Transaction / account balance data & 0.0071 & 0.0048 \\
        \bottomrule
    \end{tabular}
    \vspace{0.5em}\\
    {\scriptsize* Features are sorted by total gain; see Table \ref{tab:variables} for variable definitions}
    \label{tab:btc_importances}
\end{table}

\begin{table}[H]
    \centering
    \footnotesize
    \caption{The 50 most influential features for predicting ETH price movements using XGBoost}
    \begin{tabular}{p{0.37\textwidth} p{0.27\textwidth} R{0.1\textwidth} R{0.12\textwidth}}
        \toprule
        \textbf{Feature name*} & \textbf{Category} & \textbf{Normalised total gain} & \textbf{Normalised average gain} \\
        \midrule
        tweets\_twitter\_roberta\_pretrained\_score & NLP data & 0.0578 & 0.0440 \\
        tweets\_roberta\_finetuned\_score & NLP data & 0.0540 & 0.0372 \\
        news\_roberta\_finetuned\_score & NLP data & 0.0472 & 0.0476 \\
        indicator\_CR & Technical indicators & 0.0397 & 0.0220 \\
        reddit\_roberta\_finetuned\_score & NLP data & 0.0397 & 0.0247 \\
        indicator\_EMA & Technical indicators & 0.0359 & 0.0519 \\
        indicator\_Ichimoku\_A & Technical indicators & 0.0302 & 0.0308 \\
        indicator\_VWAP & Technical indicators & 0.0301 & 0.0355 \\
        indicator\_DCM & Technical indicators & 0.0300 & 0.0345 \\
        indicator\_BBM & Technical indicators & 0.0290 & 0.0492 \\
        indicator\_MI & Technical indicators & 0.0278 & 0.0140 \\
        total\_issues & GitHub metrics & 0.0264 & 0.0240 \\
        indicator\_WilliamsR & Technical indicators & 0.0252 & 0.0145 \\
        tweets\_bart\_mnli\_bullish\_score & NLP data & 0.0233 & 0.0122 \\
        indicator\_KAMA & Technical indicators & 0.0225 & 0.0279 \\
        indicator\_Ichimoku\_Conversion & Technical indicators & 0.0223 & 0.0253 \\
        indicator\_WMA & Technical indicators & 0.0222 & 0.0285 \\
        indicator\_CMF & Technical indicators & 0.0211 & 0.0133 \\
        indicator\_TRIX & Technical indicators & 0.0209 & 0.0480 \\
        indicator\_Stoch\_RSI & Technical indicators & 0.0193 & 0.0121 \\
        indicator\_ROC & Technical indicators & 0.0188 & 0.0199 \\
        indicator\_KCM & Technical indicators & 0.0179 & 0.0555 \\
        indicator\_FI & Technical indicators & 0.0166 & 0.0144 \\
        closed\_issues & GitHub metrics & 0.0135 & 0.0133 \\
        price\_close & Past price data & 0.0129 & 0.0097 \\
        average\_transaction\_value & Transaction / account balance data & 0.0128 & 0.0076 \\
        reddit\_accounts\_active\_48h & Numerical social media data & 0.0123 & 0.0046 \\
        indicator\_ultimate & Technical indicators & 0.0107 & 0.0076 \\
        indicator\_Ichimoku\_Base & Technical indicators & 0.0107 & 0.0165 \\
        exchange\_Coinbase\_volumefrom & Exchange volume data & 0.0097 & 0.0057 \\
        twitter\_followers & Numerical social media data & 0.0097 & 0.0039 \\
        indicator\_Ichimoku\_B & Technical indicators & 0.0092 & 0.0227 \\
        unique\_addresses\_all\_time & Transaction / account balance data & 0.0091 & 0.0060 \\
        indicator\_DCW & Technical indicators & 0.0090 & 0.0059 \\
        staking\_rate & Technical blockchain metrics & 0.0089 & 0.0099 \\
        indicator\_KST & Technical indicators & 0.0084 & 0.0204 \\
        exchange\_Kraken\_volumeto & Exchange volume data & 0.0083 & 0.0042 \\
        zero\_balance\_addresses\_all\_time & Transaction / account balance data & 0.0082 & 0.0070 \\
        indicator\_AO & Technical indicators & 0.0076 & 0.0092 \\
        indicator\_EMV & Technical indicators & 0.0075 & 0.0048 \\
        indicator\_VPT & Technical indicators & 0.0074 & 0.0066 \\
        indicator\_DPO & Technical indicators & 0.0071 & 0.0051 \\
        indicator\_Aroon\_down & Technical indicators & 0.0069 & 0.0071 \\
        total\_volume & Exchange volume data & 0.0067 & 0.0038 \\
        indicator\_CCI & Technical indicators & 0.0067 & 0.0070 \\
        indicator\_Vortex\_down & Technical indicators & 0.0065 & 0.0045 \\
        indicator\_MACD & Technical indicators & 0.0064 & 0.0128 \\
        indicator\_Stoch & Technical indicators & 0.0064 & 0.0039 \\
        USD\_volumeto & Exchange volume data & 0.0062 & 0.0029 \\
        exchange\_Kraken\_volumefrom & Exchange volume data & 0.0062 & 0.0039 \\
        \bottomrule
    \end{tabular}
    \vspace{0.5em}\\
    {\scriptsize* Features are sorted by total gain; see Table \ref{tab:variables} for variable definitions}
    \label{tab:eth_importances}
\end{table}


\section{Documentation}

\renewcommand{\thetable}{B.\arabic{table}} 
\setcounter{table}{0} 

\begin{table}[H]
	\centering
	\footnotesize
	\caption{Search ranges of the hyperparameter optimisation for the RoBERTa fine-tuning}
	\vspace{0.8em}
	\begin{tabularx}{0.5\textwidth}{p{0.24\textwidth}X}
	\toprule
	\textbf{Hyperparameter} & \textbf{Search range} \\
	\midrule
	learning rate & $5\!\times\!10^{-6}$ to 0.05 \\
	epochs & 2 to 9 \\
	batch size & 8, 16, 32, 64 \\
	warmup steps & 0 to 20 \\
	L2 regularisation parameter & 0.001 to 0.2 \\
	\bottomrule
	\end{tabularx}
	\label{tab:optuna_nlp}
\end{table}

\begin{table}[H]
	\centering
	\footnotesize
	\caption{Search ranges of the hyperparameter optimisation for the time series models}
	\vspace{0.6em}
	\begin{tabularx}{\textwidth}{p{0.1\textwidth}p{0.27\textwidth}p{0.63\textwidth}}
	\toprule
	\textbf{Model} & \textbf{Hyperparameter} & \textbf{Search range} \\
	\midrule
	Ridge & L2 regularisation parameter & $0.001$ to $100$ \\
	Regression & solver & SVD, Cholesky, LSQR, Sparse CG, SAG, SAGA \\
	\midrule
	Logistic & L2 regularisation parameter & $0.0005$ to $1000$ \\
	Regression & solver & L-BFGS, Liblinear, Newton-CG, Newton-Cholesky, SAG, SAGA \\
	\midrule
	XGBoost & number of estimators & $100$ to $1400$ \\
	& max depth & $1$ to $20$ \\
	& learning rate & $0.01$ to $0.3$ \\
	& subsampling ratio of instances & $0.5$ to $1$ \\
	& subsampling ratio of features & $0.5$ to $1$ \\
	& L1 regularisation parameter & $0.001$ to $1$ \\
	& L2 regularisation parameter & $0.001$ to $1$ \\
	& partitioning threshold (gamma) & $0$ to $1$ \\
	\midrule
	MLP & number of layers & $1$ to $4$ \\
	(FNN) & size of each layer* & $10$ to $200$ \\
	& activation function & identity (linear), logistic, hyperbolic tangent, ReLU \\
	& optimiser & L-BFGS, SGD, Adam \\
	& L2 regularisation parameter & $0.0001$ to $0.1$ \\
	& learning rate & $0.001$ to $0.1$ \\
	& scaling & none, standardisation, min-max scaling \\
	& epochs & $10$ to $1000$ \\
	& batch size & $16$, $32$, $64$, $128$ \\
	\midrule
	LSTM & number of LSTM layers & $1$ to $3$ \\
	& size of each LSTM layer* & $50$ to $300$ \\
	& number of dense layers & $0$ to $3$ \\
	& size of each dense layer* & $10$ to $150$ \\
	& activation function & hyperbolic tangent, ReLU \\
	& dropout & $0.1$ to $0.5$ \\
	& optimiser & Adam, RMSprop, SGD \\
	& learning rate & $0.0001$ to $0.1$ \\
	& scaling & none, standardisation, min-max scaling \\
	& epochs & $10$ to $200$ \\
	& batch size & $32$, $64$, $128$, $256$ \\
	\midrule
	TFT & number of LSTM layers & $1$ to $3$ \\
	& number of attention heads & $4$, $8$, $16$ \\
	& size of variable selection GRNs & $16$, $32$, $64$, $128$ \\
	& size of remaining layers** & $16$, $32$, $64$, $128$ \\
	& dropout & $0.1$ to $0.5$ \\
	& learning rate & $5\!\times\!10^{-5}$ to $0.01$ \\
	& optimiser & Adam, RMSprop, SGD, Adagrad, Ranger \\
	& gradient clipping value & $0.1$ to $1.0$ \\
	& limit\_train\_batches & $0.8$ to $1.0$ \\
	& reduce\_on\_plateau\_patience & $5$, $10$, $15$ \\
	& epochs & $1$ to $200$ \\
	& batch size & $16$, $32$, $64$, $128$, $256$ \\
	\bottomrule
	\end{tabularx}
	\scriptsize{\\\vspace{1em}* The number of neurons was tuned individually for each layer, not set uniformly for all\\\vspace{0.3em}
	** The number of neurons was set uniformly for all layers}
	\label{tab:optuna}
\end{table}

\begin{table}[H]
	\centering
	\scriptsize
	\caption{Overview and description of the Bitcoin/Ethereum features}
	\begin{tabularx}{\textwidth}{p{0.27\textwidth}p{0.12\textwidth}p{0.07\textwidth}p{0.01\textwidth}X}
		\toprule
		\textbf{Variable name} & \textbf{Source} & \textbf{Interval} & \textbf{I*} & \textbf{Description} \\
		\midrule
		price\_close & CryptoCompare & timepoint & 1 & BTC (ETH) market value in EUR calculated with the CCCAGG method (weighted average of EUR prices of 301 exchanges -- weighted by exchange volume and time since last trade) \\
		\midrule
		total\_volume & CoinGecko & 24h & 1 & Total value in EUR of BTC (ETH) that has been bought and sold on the spot market on 639 exchanges \\
		\midrule
		news\_bart\_mnli\_bullish\_score & Google News /\newline Own calculation & 24h & 0 & Average BART MNLI bullish score of news \\
		\midrule
		tweets\_bart\_mnli\_bullish\_score & Twitter /\newline  Own calculation & 24h & 0 & Average BART MNLI bullish score of Twitter posts \\
		\midrule
		reddit\_bart\_mnli\_bullish\_score & Reddit /\newline  Own calculation & 24h & 0 & Average BART MNLI bullish score of Reddit posts \\
		\midrule
		news\_twitter\_roberta\_pretrained\_score & Google News /\newline  Own calculation & 24h & 0 & Average Twitter-RoBERTa sentiment score of news \\
		\midrule
		tweets\_twitter\_roberta\_pretrained\_score & Twitter /\newline  Own calculation & 24h & 0 & Average Twitter-RoBERTa sentiment score of Twitter posts \\
		\midrule
		reddit\_twitter\_roberta\_pretrained\_score & Reddit /\newline  Own calculation & 24h & 0 & Average Twitter-RoBERTa sentiment score of Reddit posts \\
		\midrule
		news\_roberta\_finetuned\_score & Google News /\newline  Own calculation & 24h & 0 & Average Finetuned RoBERTa score of news \\
		\midrule
		tweets\_roberta\_finetuned\_score & Twitter /\newline  Own calculation & 24h & 0 & Average Finetuned RoBERTa score of Twitter posts \\
		\midrule
		reddit\_roberta\_finetuned\_score & Reddit /\newline  Own calculation & 24h & 0 & Average Finetuned RoBERTa score of Reddit posts \\
		\midrule
		news\_count & Google News & 24h & 1 & Number of news articles from CoinDesk, Cointelegraph or Decrypt for the keywords Bitcoin, BTC (Ethereum, ETH)
		\\
		\midrule
		tweet\_count & Twitter & 24h & 1 & Number of tweets containing hashtags \#bitcoin or \#btc (\#ethereum or \#eth) \\
		\midrule
		twitter\_followers & Twitter & timepoint & 2 & Count of followers of the twitter account @Bitcoin (@ethereum) \\
		\midrule
		reddit\_count & Reddit & 24h & 1 & Number of Reddit posts on r/Bitcoin (r/ethereum) \\
		\midrule
		reddit\_subscribers & Reddit & timepoint & 2 & Count of subscribers to the subreddit r/Bitcoin (r/ethereum) \\
		\midrule
		reddit\_accounts\_active\_48h & Reddit & 48h & 1 & Count of reddit accounts active on the subreddit r/Bitcoin (r/ethereum)\\
		\midrule
		forks & GitHub & timepoint & 2 & Number of forks on the bitcoin/bitcoin (ethereum/go-ethereum) GitHub repository \\
		\midrule
		stars & GitHub & timepoint & 2 & Number of stars on the GitHub repository \\
		\midrule
		subscribers & GitHub & timepoint & 2 & Number of watchers of the GitHub repository \\
		\midrule
		total\_issues & GitHub & timepoint & 2 & Number of open and closed issues of the GitHub repository \\
		\midrule
		closed\_issues & GitHub & timepoint & 2 & Number of closed issues of the GitHub repository \\
		\midrule
		pull\_requests\_merged & GitHub & timepoint & 2 & Number of merged pull requests of the GitHub repository \\
		\midrule
		pull\_request\_contributors & GitHub & timepoint & 2 & Number of pull request contributors of the GitHub repository \\
		\midrule
		additions & GitHub & 24h & 1 & Number of additions on the GitHub repository \\
		\midrule
		deletions & GitHub & 24h & 1 & Number of deletions on the GitHub repository \\
		\midrule
	\end{tabularx}
	\\\vspace*{0.7em}
	{* Order of integration (number of times the time series had to be differenced to become stationary)}
	\label{tab:variables}
\end{table}

\begin{table}[H]
	\centering
	\footnotesize
	Table \ref{tab:variables} continued\\
	\vspace{0.7em}
	\scriptsize
	\begin{tabularx}{\textwidth}{p{0.27\textwidth}p{0.12\textwidth}p{0.07\textwidth}p{0.01\textwidth}X}
		\toprule
		\textbf{Variable name} & \textbf{Source} & \textbf{Interval} & \textbf{I*} & \textbf{Description} \\
		\midrule
		commit\_count\_4\_weeks & GitHub & 4 weeks & 1 & Number of commits in the past 4 weeks on the GitHub repository \\
		\midrule
		ETH\_volumefrom (BTC\_volumefrom) & CryptoCompare & 24h & 1 & Volume of transactions from ETH to BTC (BTC to ETH) across 301 exchanges \\
		\midrule
		ETH\_volumeto (BTC\_volumeto) & CryptoCompare & 24h & 1 & Volume of transactions from BTC to ETH (ETH to BTC) across 301 exchanges \\
		\midrule
		USD\_volumefrom & CryptoCompare & 24h & 1 & Volume of transactions from USD to BTC (ETH) across 301 exchanges \\
		\midrule
		USD\_volumeto & CryptoCompare & 24h & 1 & Volume of transactions from BTC (ETH) to USD across 301 exchanges \\
		\midrule
		EUR\_volumefrom & CryptoCompare & 24h & 1 & Volume of transactions from EUR to BTC (ETH) across 301 exchanges \\
		\midrule
		EUR\_volumeto & CryptoCompare & 24h & 1 & Volume of transactions from BTC (ETH) to EUR across 301 exchanges \\
		\midrule
		exchange\_Bitfinex\_volumeto & CryptoCompare & 24h & 1 & Inflow of BTC (ETH) on Bitfinex exchange \\
		\midrule
		exchange\_Bitfinex\_volumefrom & CryptoCompare & 24h & 1 & Outflow of BTC (ETH) on Bitfinex exchange \\
		\midrule
		exchange\_Bitfinex\_volumetotal & CryptoCompare & 24h & 1 & Total BTC (ETH) cashflows on Bitfinex exchange \\
		\midrule
		exchange\_Kraken\_volumeto & CryptoCompare & 24h & 1 & Inflow of BTC (ETH) on Kraken exchange \\
		\midrule
		exchange\_Kraken\_volumefrom & CryptoCompare & 24h & 1 & Outflow of BTC (ETH) on Kraken exchange \\
		\midrule
		exchange\_Kraken\_volumetotal & CryptoCompare & 24h & 1 & Total BTC (ETH) cashflows on Kraken exchange \\
		\midrule
		exchange\_Coinbase\_volumeto & CryptoCompare & 24h & 1 & Inflow of BTC (ETH) on Coinbase exchange \\
		\midrule
		exchange\_Coinbase\_volumefrom & CryptoCompare & 24h & 1 & Outflow of BTC (ETH) on Coinbase exchange \\
		\midrule
		exchange\_Coinbase\_volumetotal & CryptoCompare & 24h & 1 & Total BTC (ETH) cashflows on Coinbase exchange \\
		\midrule
		exchange\_BTSE\_volumeto & CryptoCompare & 24h & 1 & Inflow of BTC (ETH) on BTSE exchange \\
		\midrule
		exchange\_BTSE\_volumefrom & CryptoCompare & 24h & 1 & Outflow of BTC (ETH) on BTSE exchange \\
		\midrule
		exchange\_BTSE\_volumetotal & CryptoCompare & 24h & 1 & Total BTC (ETH) cashflows on BTSE exchange \\
		\midrule
		exchange\_Binance\_volumeto & CryptoCompare & 24h & 1 & Inflow of BTC (ETH) on Binance exchange \\
		\midrule
		exchange\_Binance\_volumefrom & CryptoCompare & 24h & 1 & Outflow of BTC (ETH) on Binance exchange \\
		\midrule
		exchange\_Binance\_volumetotal & CryptoCompare & 24h & 1 & Total BTC (ETH) cashflows on Binance exchange \\
		\midrule
		zero\_balance\_addresses\_all\_time & IntoTheBlock & timepoint & 2 & Amount of BTC (ETH) addresses that have always had zero balance since inception \\
		\midrule
		unique\_addresses\_all\_time & IntoTheBlock & timepoint & 2 & Amount of BTC (ETH) addresses that executed at least one transaction since inception \\
		\midrule
		new\_addresses & IntoTheBlock & 24h & 1 & Amount of new BTC (ETH) addresses created \\
		\midrule
		active\_addresses & IntoTheBlock & 24h & 1 & Amount of BTC (ETH) addresses that executed at least one transaction \\
		\midrule
		transaction\_count & IntoTheBlock & 24h & 1 & Number of valid transactions on the BTC (ETH) blockchain \\
		\midrule
		large\_transaction\_count & IntoTheBlock & 24h & 1 & Number of valid transactions greater than 100,000 USD on the BTC (ETH) blockchain \\
		\midrule
		average\_transaction\_value & IntoTheBlock & 24h & 1 & Average transaction value on the BTC (ETH) blockchain in BTC (ETH) \\
		\midrule
		hashrate (only Bitcoin) & IntoTheBlock & 24h & 1 & Number of terahashes per second the BTC network is performing \\
		\midrule
		difficulty (only Bitcoin) & IntoTheBlock & 24h & 1 & Mean difficulty of finding a hash that meets the protocol-designated requirement (difficulty is adjusted every 2016 blocks so that the average time between each block remains $\sim$10 minutes) \\
		\midrule
		block\_time (only Bitcoin) & IntoTheBlock & 24h & 1 & Average time in seconds it takes miners to verify transactions within one block on the BTC network \\
		\midrule
	\end{tabularx}
\end{table}

\begin{table}[H]
	\centering
	\footnotesize
	Table \ref{tab:variables} continued\\
	\vspace{0.7em}
	\scriptsize
	\begin{tabularx}{\textwidth}{p{0.27\textwidth}p{0.12\textwidth}p{0.07\textwidth}p{0.01\textwidth}X}
		\toprule
		\textbf{Variable name} & \textbf{Source} & \textbf{Interval} & \textbf{I*} & \textbf{Description} \\
		\midrule
		block\_size & IntoTheBlock & 24h & 1 & Average block size in bytes on the BTC (ETH) blockchain \\
		\midrule
		current\_supply & IntoTheBlock & timepoint & 2 & Sum of all BTC (ETH) issued on the BTC (ETH) ledger \\
		\midrule
		staking\_rate (only Ethereum) & Attestant & 24h & 2 & ETH staking yield (1 year ROI of staking ETH) offered by Attestant \\
		\midrule
		balance\_distribution\_from\_0.0 \newline \_totalVolume & IntoTheBlock & timepoint & 1 & Total amount of BTC (ETH) held by addresses with a balance between 0 and 0.001 BTC (ETH) \\
		\midrule
		balance\_distribution\_from\_0.001 \newline \_totalVolume & IntoTheBlock & timepoint & 1 & Total amount of BTC (ETH) held by addresses with a balance between 0.001 and 0.01 BTC (ETH) \\
		\midrule
		balance\_distribution\_from\_0.01 \newline \_totalVolume & IntoTheBlock & timepoint & 1 & Total amount of BTC (ETH) held by addresses with a balance between 0.01 and 0.1 BTC (ETH) \\
		\midrule
		balance\_distribution\_from\_0.1 \newline \_totalVolume & IntoTheBlock & timepoint & 1 & Total amount of BTC (ETH) held by addresses with a balance between 0.1 and 1 BTC (ETH) \\
		\midrule
		balance\_distribution\_from\_1.0 \newline \_totalVolume & IntoTheBlock & timepoint & 1 & Total amount of BTC (ETH) held by addresses with a balance between 1 and 10 BTC (ETH) \\
		\midrule
		balance\_distribution\_from\_10.0 \newline \_totalVolume & IntoTheBlock & timepoint & 1 & Total amount of BTC held by addresses with a balance between 10 and 100 BTC \\
		\midrule
		balance\_distribution\_from\_100.0 \newline \_totalVolume & IntoTheBlock & timepoint & 1 & Total amount of BTC (ETH) held by addresses with a balance between 100 and 1000 BTC (ETH) \\
		\midrule
		balance\_distribution\_from\_1000.0 \newline \_totalVolume & IntoTheBlock & timepoint & 1 & Total amount of addresses with a balance between 1000 and 10000 BTC (ETH) \\
		\midrule
		balance\_distribution\_from\_10000.0 \newline \_totalVolume & IntoTheBlock & timepoint & 1 & Total amount of addresses with a balance between 10000 and 100000 BTC (ETH) \\
		\midrule
		balance\_distribution\_from\_100000.0 \newline \_totalVolume & IntoTheBlock & timepoint & 1 & Total amount of addresses with a balance above 100000 BTC (ETH) \\
		\midrule
		balance\_distribution\_from\_0.0 \newline \_addressesCount & IntoTheBlock & timepoint & 1 & Total amount of addresses with a balance between 0 and 0.001 BTC (ETH) \\
		\midrule
		balance\_distribution\_from\_0.001 \newline \_addressesCount & IntoTheBlock & timepoint & 1 & Total amount of addresses with a balance between 0.001 and 0.01 BTC (ETH) \\
		\midrule
		balance\_distribution\_from\_0.01 \newline \_addressesCount & IntoTheBlock & timepoint & 1 & Total amount of addresses with a balance between 0.01 and 0.1 BTC (ETH) \\
		\midrule
		balance\_distribution\_from\_0.1 \newline \_addressesCount & IntoTheBlock & timepoint & 1 & Total amount of addresses with a balance between 0.1 and 1 BTC (ETH) \\
		\midrule
		balance\_distribution\_from\_1.0 \newline \_addressesCount & IntoTheBlock & timepoint & 1 & Total amount of addresses with a balance between 1 and 10 BTC (ETH) \\
		\midrule
		balance\_distribution\_from\_10.0 \newline \_addressesCount & IntoTheBlock & timepoint & 1 & Total amount of addresses with a balance between 10 and 100 BTC (ETH) \\
		\midrule
		balance\_distribution\_from\_100.0 \newline \_addressesCount & IntoTheBlock & timepoint & 1 & Total amount of addresses with a balance between 100 and 1000 BTC (ETH) \\
		\midrule
		balance\_distribution\_from\_1000.0 \newline \_addressesCount & IntoTheBlock & timepoint & 1 & Total amount of addresses with a balance between 1000 and 10000 BTC (ETH) \\
		\midrule
		balance\_distribution\_from\_10000.0 \newline \_addressesCount & IntoTheBlock & timepoint & 1 & Total amount of addresses with a balance between 10000 and 100000 BTC (ETH) \\
		\midrule
		balance\_distribution\_from\_100000.0 \newline \_addressesCount & IntoTheBlock & timepoint & 1 & Total amount of addresses with a balance above 100000 BTC (ETH) \\
		\midrule
		index\_MVDA\_close & CryptoCompare & timepoint & 1 & Close value of the MarketVector Digital Assets 100 (MVDA) index tracking the 100 largest digital assets \\
		\midrule
		index\_BVIN\_close & CryptoCompare & timepoint & 1 & Close value of the CryptoCompare Bitcoin Volatility (BVIN) index tracking BTC implied volatility using options data from Deribit \\
		\midrule
		gtrends\_bitcoin\_relative\_change & Google Trends & 24h & 1 & Percentage change in amount of Google searches of the keyword ``bitcoin'' \\
		\midrule
		gtrends\_ethereum\_relative\_change & Google Trends & 24h & 1 & Percentage change in amount of Google searches of the keyword ``ethereum'' \\
		\midrule
	\end{tabularx}
\end{table}

\begin{table}[H]
	\centering
	\footnotesize
	Table \ref{tab:variables} continued\\
	\vspace{0.7em}
	\scriptsize
	\begin{tabularx}{\textwidth}{p{0.27\textwidth}p{0.12\textwidth}p{0.07\textwidth}p{0.01\textwidth}X}
		\toprule
		\textbf{Variable name} & \textbf{Source} & \textbf{Interval} & \textbf{I*} & \textbf{Description} \\
		\midrule
		gtrends\_cryptocurrency\_relative\_change & Google Trends & 24h & 1 & Percentage change in amount of Google searches of the keyword ``cryptocurrency'' \\
		\midrule
		gtrends\_blockchain\_relative\_change & Google Trends & 24h & 1 & Percentage change in amount of Google searches of the keyword ``blockchain'' \\
		\midrule
		gtrends\_investing\_relative\_change & Google Trends & 24h & 1 & Percentage change in amount of Google searches of the keyword ``investing'' \\
		\midrule
		sp500\_price & Yahoo Finance & timepoint & 1 & Close value of the S\&P 500 index \\
		\midrule
		sp500\_volume & Yahoo Finance & 24h & 1 & Volume of the S\&P 500 index \\
		\midrule
		vix & Yahoo Finance & timepoint & 1 & Close value of the CBOE Volatility Index (VIX) tracking the S\&P 500 implied volatility \\
		\midrule
		gold\_usd\_price & Yahoo Finance & timepoint & 1 & Close value of the COMEX gold future (GC) \\
		\midrule
		indicator\_AO & Own calculation & timepoint & 1 & Awesome Oscillator (AO) - Measures market momentum to capture the potential change in trend \\
		\midrule
		indicator\_KAMA & Own calculation & timepoint & 1 & Kaufman's Adaptive Moving Average (KAMA) - A moving average that adjusts its length based on market volatility \\
		\midrule
		indicator\_PPO & Own calculation & timepoint & 1 & Percentage Price Oscillator (PPO) - Measures the difference between two moving averages as a percentage of the larger moving average \\
		\midrule
		indicator\_PVO & Own calculation & timepoint & 1 & Percentage Volume Oscillator (PVO) - Like PPO but for volume, it measures the difference between two volume moving averages \\
		\midrule
		indicator\_ROC & Own calculation & timepoint & 1 & Rate of Change (ROC) - Measures the percentage change in price from one period to the next \\
		\midrule
		indicator\_RSI & Own calculation & timepoint & 1 & Relative Strength Index (RSI) - Measures the speed and change of price movements and indicates overbought or oversold conditions \\
		\midrule
		indicator\_Stoch\_RSI & Own calculation & timepoint & 1 & Stochastic RSI - Combines stochastic oscillator and RSI to measure the RSI relative to its high-low range \\
		\midrule
		indicator\_Stoch & Own calculation & timepoint & 1 & Stochastic Oscillator - Compares a closing price to its price range over a specific time period \\
		\midrule
		indicator\_TSI & Own calculation & timepoint & 1 & True Strength Index (TSI) - Measures the momentum of price movements \\
		\midrule
		indicator\_ultimate & Own calculation & timepoint & 1 & Ultimate Oscillator - Combines short, medium, and long-term price action into one oscillator to avoid false divergences \\
		\midrule
		indicator\_WilliamsR & Own calculation & timepoint & 1 & Williams \%R - A momentum indicator that measures overbought/oversold levels \\
		\midrule
		indicator\_ADI & Own calculation & timepoint & 1 & Accumulation/Distribution Index (ADI) - Measures the cumulative flow of money into and out of a security \\
		\midrule
		indicator\_CMF & Own calculation & timepoint & 1 & Chaikin Money Flow (CMF) - Measures the amount of Money Flow Volume over a specific period \\
		\midrule
		indicator\_EMV & Own calculation & timepoint & 1 & Ease of Movement (EMV) - Relates volume and price change to show how much volume is needed to move prices \\
		\midrule
		indicator\_FI & Own calculation & timepoint & 1 & Force Index (FI) - Measures the buying or selling pressure over a specific period \\
		\midrule
		indicator\_MFI & Own calculation & timepoint & 1 & Money Flow Index (MFI) - A volume-weighted version of RSI that shows price strength \\
		\midrule
		indicator\_NVI & Own calculation & timepoint & 1 & Negative Volume Index (NVI) - Focuses on days where the volume decreases from the previous day \\
		\midrule
		indicator\_OBV & Own calculation & timepoint & 1 & On-Balance Volume (OBV) - Relates volume to price change \\
		\midrule
		indicator\_VPT & Own calculation & timepoint & 1 & Volume Price Trend (VPT) - Combines price and volume to show the direction of price trend \\
		\midrule
	\end{tabularx}
\end{table}

\begin{table}[H]
	\centering
	\footnotesize
	Table \ref{tab:variables} continued\\
	\vspace{0.7em}
	\scriptsize
	\begin{tabularx}{\textwidth}{p{0.27\textwidth}p{0.12\textwidth}p{0.07\textwidth}p{0.01\textwidth}X}
		\toprule
		\textbf{Variable name} & \textbf{Source} & \textbf{Interval} & \textbf{I*} & \textbf{Description} \\
		\midrule
		indicator\_VWAP & Own calculation & timepoint & 1 & Volume Weighted Average Price (VWAP) - The average price weighted by volume \\
		\midrule
		indicator\_BBM & Own calculation & timepoint & 1 & Bollinger Middle Band - The middle band in the Bollinger Bands, which is a simple moving average \\
		\midrule
		indicator\_BBW & Own calculation & timepoint & 1 & Bollinger Bandwidth - The width of the Bollinger Bands \\
		\midrule
		indicator\_DCM & Own calculation & timepoint & 1 & Donchian Channel Middle Band - The average of the Donchian high and low bands \\
		\midrule
		indicator\_DCW & Own calculation & timepoint & 1 & Donchian Channel Width - The width of the Donchian Bands \\
		\midrule
		indicator\_KCM & Own calculation & timepoint & 1 & Keltner Channel Middle Band - The average of the Keltner high and low bands \\
		\midrule
		indicator\_KCW & Own calculation & timepoint & 1 & Keltner Channel Width - The width of the Keltner Bands \\
		\midrule
		indicator\_UI & Own calculation & timepoint & 1 & Ulcer Index (UI) - Measures downside risk in terms of price declines \\
		\midrule
		indicator\_Aroon\_down & Own calculation & timepoint & 1 & Aroon Down - Identifies the number of days since a 25-day low \\
		\midrule
		indicator\_Aroon\_up & Own calculation & timepoint & 1 & Aroon Up - Identifies the number of days since a 25-day high \\
		\midrule
		indicator\_CCI & Own calculation & timepoint & 1 & Commodity Channel Index (CCI) - Measures the difference between a security's price change and its average price change. \\
		\midrule
		indicator\_DPO & Own calculation & timepoint & 1 & Detrended Price Oscillator (DPO) - Removes trend from price. \\
		\midrule
		indicator\_EMA & Own calculation & timepoint & 1 & Exponential Moving Average (EMA) - A moving average that gives more weight to recent prices. \\
		\midrule
		indicator\_Ichimoku\_A,\newline indicator\_Ichimoku\_B, \newline indicator\_Ichimoku\_Base, \newline indicator\_Ichimoku\_Conversion & Own calculation & timepoint & 1 & Ichimoku Cloud - A collection of technical indicators that show support and resistance levels, as well as momentum and trend direction \\
		\midrule
		indicator\_KST & Own calculation & timepoint & 1 & Know Sure Thing (KST) - A momentum oscillator based on the smoothed rate-of-change for four different time frames \\
		\midrule
		indicator\_MACD & Own calculation & timepoint & 1 & Moving Average Convergence Divergence (MACD) - Shows the relationship between two moving averages of a security's price \\
		\midrule
		indicator\_MACD\_Signal & Own calculation & timepoint & 1 & MACD Signal - A signal line for the MACD \\
		\midrule
		indicator\_MI & Own calculation & timepoint & 1 & Mass Index (MI) - Measures the volatility of price changes \\
		\midrule
		indicator\_TRIX & Own calculation & timepoint & 1 & TRIX - Shows the percent rate of change of a triple exponentially smoothed moving average \\
		\midrule
		indicator\_Vortex\_down & Own calculation & timepoint & 1 & Vortex Indicator - Identifies the start of a new trend or the continuation of a current trend \\
		\midrule
		indicator\_Vortex\_up & Own calculation & timepoint & 1 & Vortex Indicator - Identifies the start of a new trend or the continuation of a current trend \\
		\midrule
		indicator\_WMA & Own calculation & timepoint & 1 & Weighted Moving Average (WMA) - A moving average where more recent prices are given more weight \\
		\midrule
		indicator\_CR & Own calculation & timepoint & 1 & Cumulative Return (CR) - Measures the total return of a stock over a set period \\
		\midrule
		indicator\_PSAR\_down,\newline indicator\_PSAR\_up & Own calculation & timepoint & 0 & Parabolic stop and reverse - `down' (providing exit points) and `up' (providing entry points)\\
		\bottomrule
	\end{tabularx}
\end{table}

\clearpage


\bibliographystyle{cas-model2-names}
\bibliography{references}


\end{document}